\documentclass[12pt]{article}
\usepackage{caption}
\usepackage{scicite}
\usepackage{amsmath}
\usepackage{amsfonts} 
\usepackage{breqn}
\usepackage{array}
\usepackage{graphicx}
\usepackage{adjustbox}
\usepackage{hhline} 
\usepackage{hyperref}
\newcolumntype{M}[1]{>{\centering\arraybackslash}m{#1}}

\usepackage{times}

\DeclareMathOperator{\sign}{sign}

\makeatletter
\let\caption@ORIGINAL\caption
\def\caption{%
  \@ifstar
    {\captionlabelfalse          
     \addtocounter{\@captype}{-1}
     \renewcommand\addcontentsline[3]{}
     \caption@ORIGINAL[]}%
    {\caption@ORIGINAL}}
\makeatother
\newenvironment{sciabstract}{%
\begin{quote} \bf}
{\end{quote}}

\newcounter{lastnote}

\title{Multi-dimensional structure of {\it C. elegans\/} thermal learning}

\author
{Ahmed Roman$^{1}$, Konstantine Palanski$^{2}$, Ilya Nemenman$^{1,3,4}$, William S Ryu$^{5,6,\ast}$\\
\\
\normalsize{$^{1}$Department of Physics, Emory University, Atlanta, USA}\\
\normalsize{$^{2}$ANTIBODY Healthcare Communications, Toronto, Canada}\\
\normalsize{$^{3}$Department of Biology, Emory University, Atlanta, USA}\\
\normalsize{$^{4}$Initiative in Theory and Modeling of Living Systems, Emory University, Atlanta, USA}\\
\normalsize{$^{5}$Department of Physics, University of Toronto, Toronto, Canada}\\
\normalsize{$^{6}$The Donnelly Centre, University of Toronto, Toronto, Canada}\\
\\
\normalsize{$^\ast$Corresponding author; E-mail: willryu@gmail.com}
}

\date{}

\begin{document} 


\baselineskip24pt


\maketitle 


\begin{sciabstract}

Quantitative models of associative learning that explain behavior of real animals with high precision have turned out very difficult to construct. We do this in the context of the dynamics of the thermal preference of {\em C.~elegans}. For this, we quantify {\em C.~elegans} thermotaxis in response to various conditioning parameters, genetic perturbations, and operant behavior using a fast, high-throughput microfluidic droplet assay. We then model this data comprehensively, within a new, biologically interpretable, multi-modal framework. We discover that the dynamics of thermal preference are described by two independent contributions and require a model with at least four dynamical variables. One pathway positively associates the experienced temperature independently of food and the other negatively associates to the temperature when food is absent.

\end{sciabstract}


\section*{Introduction}

Animals modify their behavior through learning. Conditioning or associative learning, that is, making associations between various stimuli is one of the best studied such process \cite{Pavlov-27,Niv:2009dw,Vogel2004,Houwer2001,Krypotos2015}. Nonetheless models of conditioning, quantitatively accounting for intricacies of behavior are rare \cite{Zhao2021,Pamir2014,Faghihi2017,Otto2013}.  

In conditioning, a reward or punishment from an unconditionally appetitive or noxious stimulus (US, or reinforcement) teaches the animal to associate it with conditioned stimuli (CS, or cue), which predict the US. Conditioning is usually modeled within the Rescorla-Wagner (RW) \cite{RescorlaWagner} and the reinforcement learning (RL) frameworks \cite{RescorlaWagner,Dayan:2001uk,Dayan:2008fb,Niv:2009dw,Vogel2004}, where an error an animal makes in predicting a US from CS cues changes the strength of the CS-US associations, decreasing future errors. If the conditioning process is slow, so that many predictive cue-reward pairs happen, and the CSs appear one at a time, then the RW model becomes (see \nameref{Section:learning} for a derivation):
\begin{equation}
    \tau \frac{dV_i}{dt}= f_i(t)\left(\lambda_0f_{\lambda|i}(t)-V_i(t)\right).
    \label{eq:rl}
\end{equation}
Here $V_i$ is the association strength between the $i$th CS and the US in some arbitrary units set by $\lambda_0$, the salience of the US. Further,  $\tau$ is the time scale of learning, and $f_i$ and $f_{\lambda|i}$,  both in the range of $[0,1]$, are the frequency of the $i$th CS and the US conditional on it, respectively.  The frequencies may depend on the animal's behavior, which, in turn, is controlled by the  association strengths $\{V_i\}$. 

Such conditioning models are simplified and often lack the complexity to serve as {\em quantitative} models of real life animal behavior \cite{Niv:2009dw}. Even worse, they fail to account for some {\em qualitative} features of conditioning, including (i) extinguishing an association (not merely unlearning) and subsequent frequent spontaneous recovery \cite{Dayan:2008fb,Niv:2009dw,Dunsmoor:2015ij}, (ii) existence of multiple reinforcement systems (some potentially associated with habitual and not reward-driven actions), outputs of which integrate into behavior \cite{Daw:2005ds,Niv:2009dw}, (iii) asymmetric responses to appetitive vs.\ aversive cues \cite{Dayan:2008fb} and to conditioned association vs.\ conditioned inhibition (in the latter, a CS predicts absence of the US) \cite{Niv:2009dw}, (iv) generalization among similar, but distinct cues \cite{Niv:2009dw}, and so on.  The weakness of models is often due to the difficulty in designing informative experiments: measuring behavior with high precision, quickly, and for long duration is nontrivial; behavior itself may modify the conditioning contingency; behavior is noisy, often discrete, and hence is not a reliable readout of a CS-US association; and biology of different reinforcement pathways and mechanisms of their integration are unclear. 

To generate more useful data, and to incorporate them into more accurate models of an animal's behavior during learning, we turn to {\em C.~elegans} as a {\em quantitative} model system. The worm---one of the simplest organisms  exhibiting conditioning---associates the presence of food with various environmental signals, such as salts \cite{saeki_plasticity_2001}, odors \cite{zhang_pathogenic_2005} and temperature \cite{Hedgecock75, Mori95}, modifying  complex, but  measurable behaviors (chemo- and thermotaxis \cite{Ward_1973, Hedgecock75,Biron06,Amano2011}). It is a particularly good model system for studying associative learning because of its simple nervous system \cite{white_structure_1986}, short life cycle, and our ability to control the environment and accurately measure its behavior over long periods of time (hours). In addition, a number of genes affecting learning in {\em C.~elegans} have been identified \cite{hobert_behavioral_2003}, most notably those in the insulin-like signaling pathway \cite{stein_intersection_2012}.

{\em C.~elegans} thermotaxis is a well-studied behavior affected by conditioning: when placed on a thermal gradient, {\em C.~elegans} taxes to their cultivation temperature \cite{Hedgecock75,Mori95,Ryu02} and, when near this thermal preference, it performs isothermal tracking \cite{Hedgecock75, Mori95, Ryu02}. A new thermal preference is acquired when worms are placed at a different cultivation temperature \cite{Hedgecock75, Mori95, Biron06}. The thermal preference and its temporal dynamics can be measured by tracking worm behavior on thermal gradients \cite{Ryu02, Clark06, ramot_thermotaxis_2008, Chi07}. The thermal preference is established asymmetrically between high and low (above or below $\sim20^\circ$C) temperature conditions \cite{Ryu02,Yamada03}. The preferred temperature may depend on the starvation state of the worm \cite{Hedgecock75} and on the steepness of the temperature gradient \cite{Jurado2010, ramot_thermotaxis_2008}.  Further, the dynamics of the preference is relatively fast (from tens of minutes to a few hours) \cite{Biron06, ramot_thermotaxis_2008}. This makes designing experiments to assay the dynamics of associations difficult since the preferences can change faster than the worms move in gradients to reveal them. As a result, even some of the most basic questions about {\em C.~elegans} associative learning remain unclear, such as the relative importance of food to the establishment of the preference, precluding accurate mathematical models of the process.

Here we designed a microfluidic assay to monitor the thermotactic preference of individual worms, with the precision and the temporal resolution sufficient to track its dynamics. We showed that the simple model, Eq.~(\ref{eq:rl}), cannot precisely fit such data. Thus we developed, and then experimentally verified, a more complicated picture of the dynamics of conditional associations in {\em C.~elegans}. We identified multiple pathways affecting thermotaxis: habituation to the experienced temperature, and avoidance of temperature, at which no food was collected. Using worms with mutations in the insulin signaling pathway---whose behavior the model predicts with quantitative accuracy---we isolated contributions of these independent pathways to the behavior. We argue that the developed model solves a variety of long-standing conceptual problems in the field of animal associative learning. We further suggest that such multi-pathway organization of the conditional response dynamics may be optimal for food search in dynamical environments. 

\section*{Results}

\subsection*{Measuring thermal preference}
\begin{figure}[h!]
\includegraphics[width=10cm]{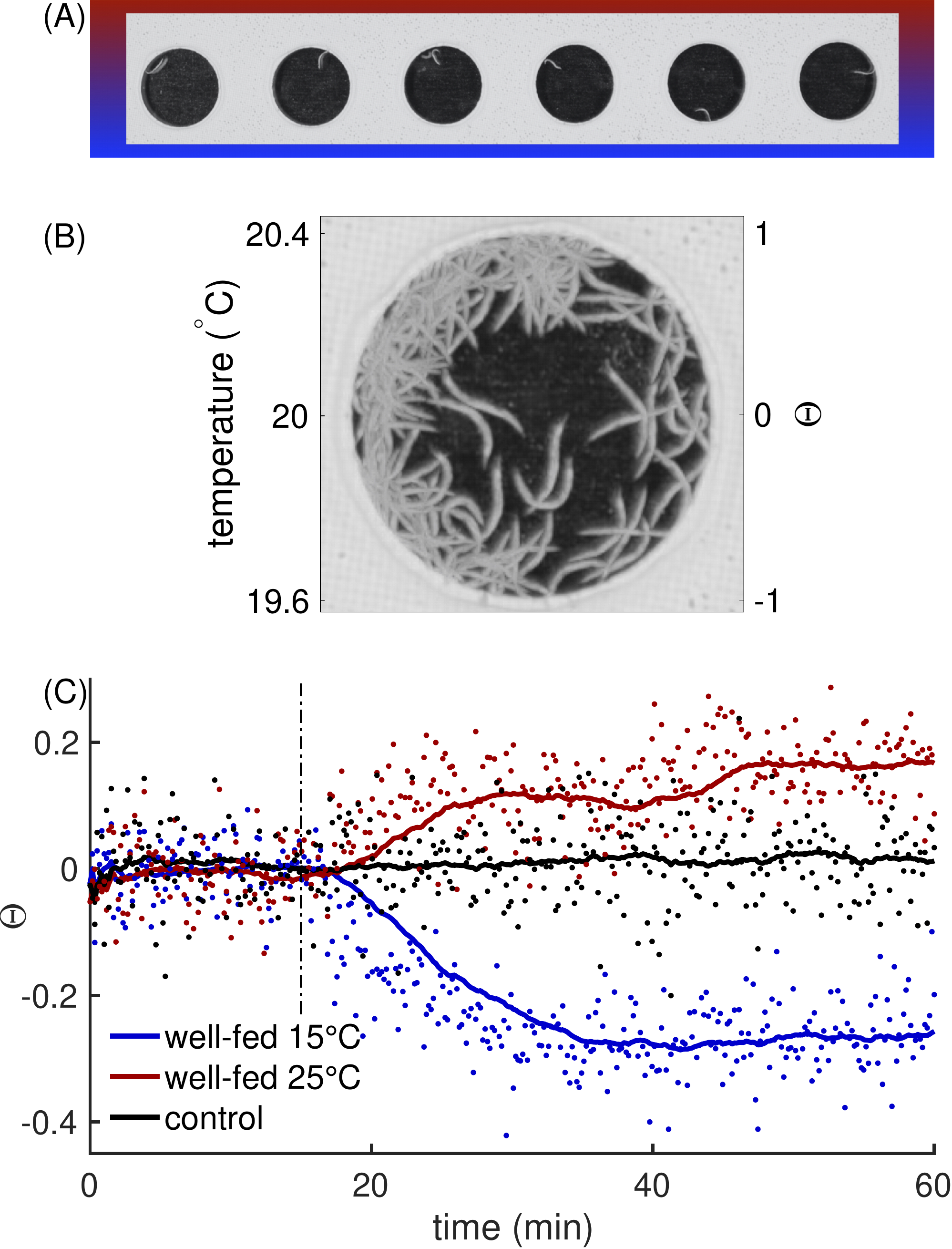} 
\caption{\textbf{Droplet thermotaxis assay}. (\textbf{A}) An array of PTFE constrained droplets with single {\em C.~elegans} placed in a thermal gradient. (\textbf{B}) Multiple exposures of a single droplet (150 frames captured at 6 frames per minute) superimposed on each other. The thermotaxis index ($\Theta$) is the average position of the worm in the droplet along the thermal gradient. (\textbf{C}) Thermotactic response of N2 with a cryophilic and thermophilic preferences (blue and red) and N2 control with no gradient (black). The vertical dashed-dotted line indicates the onset of the temperature gradient in the droplet for the biased worms.}
\label{fig:schematic}
\end{figure}

We rear N2 wild type worms at 15$^\circ$C and 25$^\circ$C, hereafter called cold and warm worms, cf.~\nameref{Section:strains}. Individual worms are then placed in each droplet (4 mm dia.) on a microscope slide with a six droplet array Fig.~\ref{fig:schematic}, cf.~\nameref{Section:droplets}. The slide is sealed, and the swimming patterns of each worm are quantified for up to 4 hours. We first acclimatize the worms for 15 min, and then turn on the thermal gradient of 1$^\circ$C/cm (from 19.6$^\circ$ to 20.4$^\circ$ across the droplet width). Since the droplet is small and can be traversed by a worm in seconds, the thermotactic bias in the worm’s position can be measured in less than a minute, allowing the quantification of its dynamics. Each worm’s position along the gradient in the droplet is rescaled to take values between -1 (cold edge) and +1 (warm edge), and the thermotactic index, $\Theta$, is determined as the mean position over some period of time (usually tens of seconds, cf.~\nameref{Section:data}). The average $\Theta$ for about 110 worms reared at 15$^\circ$C and 25$^\circ$C are shown in Fig.~\ref{fig:schematic}a  (blue and red, respectively). The cold (warm) animals initially show strong preference to the cold (warm) side, $\Theta\approx \pm 0.25$, respectively. For scale, note that the worms move constantly, and if they were to explore just one side of the droplet uniformly---a very strong bias---the thermotactic index would be $1/3$. 

\subsection*{Dynamics of thermal preference}
\begin{figure}[h!]
\includegraphics[width=10cm]{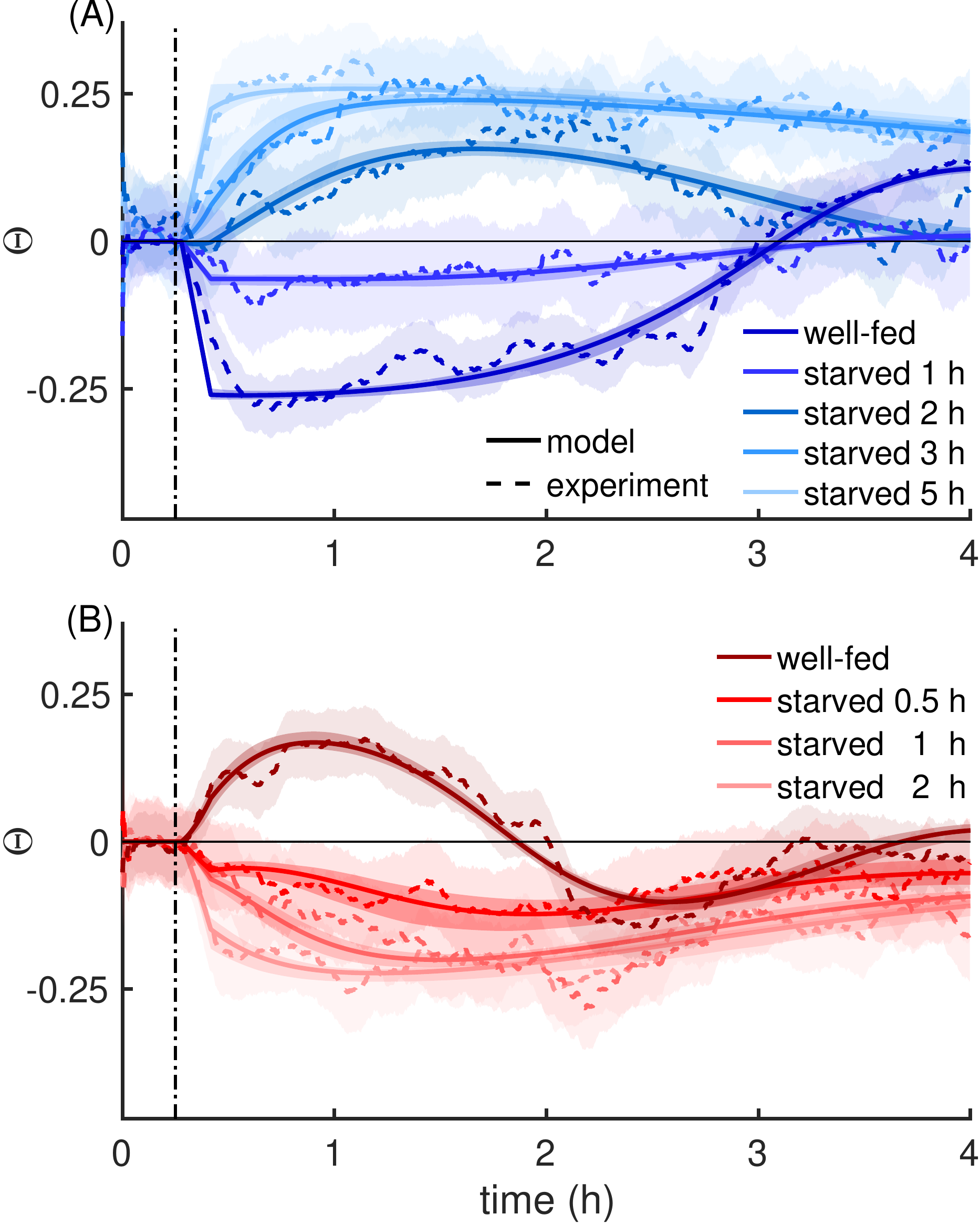} 
\caption{\textbf{Dynamics of thermal memory for wild type worms.} Thermotactic index of cold (\textbf{A}) and warm (\textbf{B}) worms with different duration of starvation is shown by different shades of blue (red). The vertical dashed-dotted line indicates the onset of the temperature gradient. Immediate preference of the colder (warmer) side by the non-starved cold (warm)  reverts to the preference of the warmer (colder) side as the worm spends more time in the droplet. The initial preference also weakens and reverses to the avoidance when the worm is subjected to long starvation before the droplet assay. Data is shown as dashed lines, and model fits are depicted with solid lines. All curves with different pre-assay starvation durations for the same rearing temperature share the same fitting parameters. For presentation purposes only (but not during fitting), both data and model are filtered, cf.~\nameref{Section:data}. Error bands on the data are the 16.5 to 83.5  percentiles. Error bands for the models are obtained using bootstrapping, cf.~\nameref{Section:bands}.}
\label{fig:wildtype}
\end{figure}

After the thermal gradient is turned on, cold (warm) worms perform thermotaxis to the cold (warm) side of the droplet, presumably expecting to find food there, cf.~Fig.~\ref{fig:schematic}(C) and Fig.~\ref{fig:wildtype}. However, the droplet has no food, and the strength of the bias decreases with time, changing sign in about 3h (2h), Fig.~\ref{fig:wildtype}. The warm worms then return back to zero bias by 3.5h in the droplet, while the cold worms remain warm-biased for the remaining time. This observation of the zero crossing in the thermal preference is crucial: the simple dynamical system,  Eq.~(\ref{eq:rl}), cannot oscillate autonomously, cf.~\nameref{Section:learning}. Thus the model in Eq.~(\ref{eq:rl}) is incomplete. Specifically, to model the observed thermotactic index dynamics, we need at least two interacting dynamical variables in a model, with two distinct time scales.

The dynamics in Fig.~\ref{fig:wildtype} proceed autonomously, with no food in the droplet, eventually developing {\em avoidance} of an initially attractive temperature. A possible explanation is a faster decay of the association on the initially attractive side than on the opposite side. However, the two sides are less than a degree apart, precluding large differences in time scales. We thus explain the avoidance by assuming that at least one of the dynamical variables describing the thermotactic index is reinforced by the {\em absence} of food, encoding avoidance of (rather than preference to) certain temperatures.

\begin{figure}[t!]
\includegraphics[width=10cm]{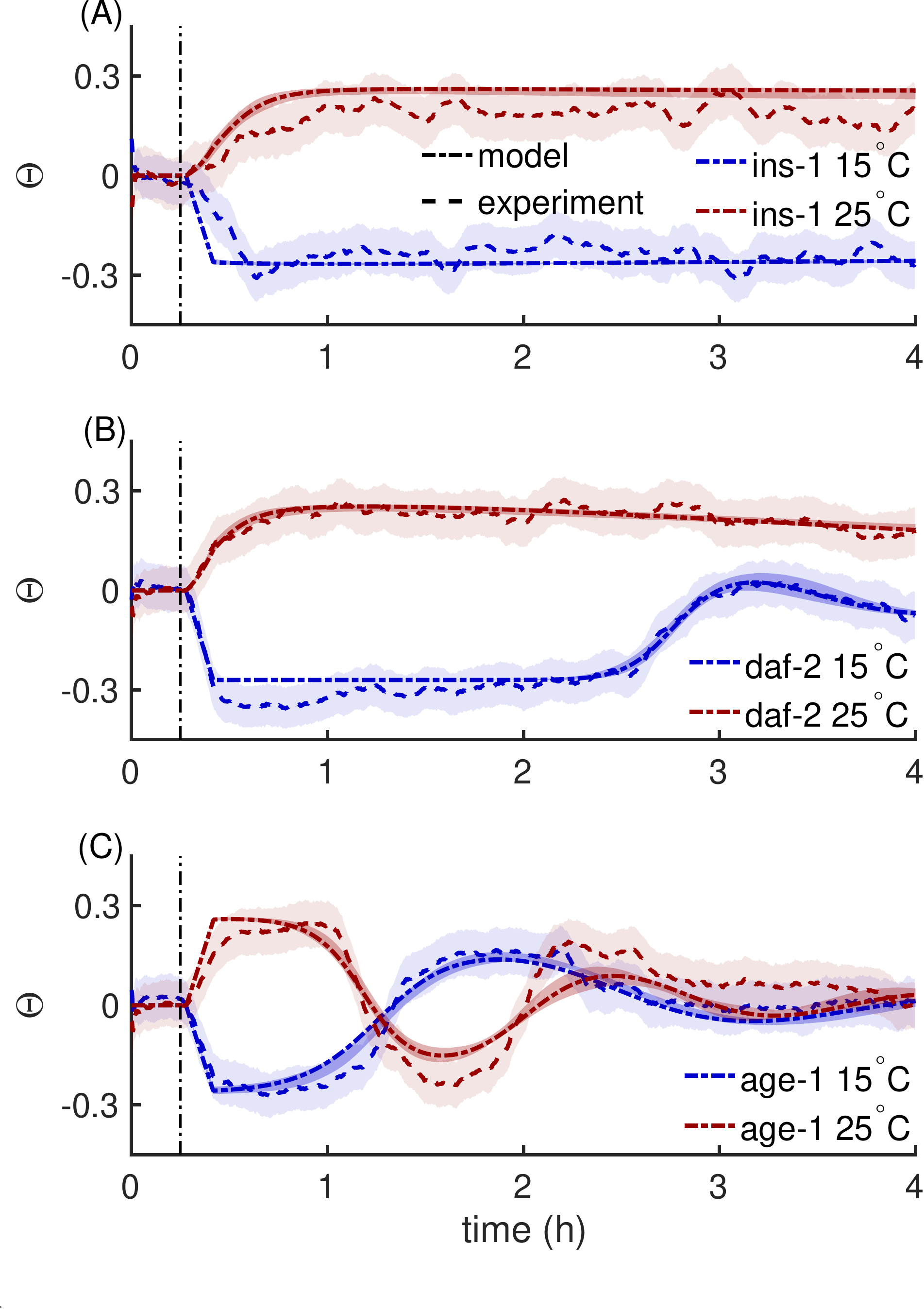}
\caption{\textbf{Dynamics of thermal memory for mutants.} Thermotactic index of the mutant worms {\it ins\/}-1 (\textbf{A}), {\it daf\/}-2 (\textbf{B})  and {\it age\/}-1 (\textbf{C})  reared at $15^\circ$C (blue) and $25^\circ$C (red). Worms are well fed before assaying the thermotactic response.  The vertical dashed-dotted line indicates the onset of the temperature gradient. Plotting conventions are as in Fig.~\ref{fig:wildtype}.}
\label{fig:mutants}     
\end{figure}

To further understand the nature of the dynamical variables involved in the thermotactic preferences, we focus on the {\em ins-1(nr2091)} strain with a mutation of an insulin-like peptide \cite{Pierce_2001}. This mutant has a more persistent thermal preference \cite{Kodama06} and shows a defective negative association in odor and salt learning \cite{Lin_2010,Tomioka2006}, but presents normal starvation behavior \cite{Kodama06}. We find, cf.~Fig.~\ref{fig:mutants}A, that, while {\em ins-1} worms reared at either 15- or 25$^\circ$C initially show the same cryophilic and thermophilic preference as the N2 wild type, the preference is sustained for the duration of the experiment. This persistence  makes it possible to interpret these data in the context of models similar to Eq.~(\ref{eq:rl}), but only if the parts of the thermal memory not affected by the {\em ins-1} mutation do not decay with time (unlikely for any non-reinforced association) or are reinforced even without food. The latter option suggests that one of the dynamical variables in model of the thermotactic index is likely to be habituation to the current temperature, rather than food-temperature association. 

In summary, zero crossing in the N2 data and absence of preference degradation in the {\em ins-1} mutant collectively suggest that an effective model must include, at least, two dynamical processes: habituation to the current temperature and avoidance of temperature when no food has been observed. Crucially, since {\em ins-1} worms exhibit no avoidance with no effect on the initial positive association, these two signals must be mediated by distinct biological pathways.

\subsection*{Constructing a model of thermal preference dynamics}
To build a mathematical model of the conditioning, we explore the dimensionality of the behavioral dynamics using tools from dynamical systems theory. First, we consider the delayed embedding representation of worm trajectories \cite{Takens81}, which bounds the number of dynamical variables by, at least, 3, cf.~\nameref{Section:dimensionality}. Second, our data sets include worms that were starved after initial rearing and before being assayed, cf.~Fig.~\ref{fig:wildtype}. Crucially, for some starvation durations, $\Theta$ does not approach its saturating values. In this linear regime, multiple time scales of oscillations suggest that at least four dynamical variables contribute to the conditioning dynamics  cf.~\nameref{Section:dimensionality}.

In other words, we seek to describe the dynamics of $\Theta$ as a combination of habituation to the current temperature, its avoidance if not reinforced by food, and two additional variables. To shed light on the latter,  we note that the worms are reared and assayed at different temperatures. Thus, for example, initial preference to the cold side at 19.6$^\circ$C is a {\em generalization}  \cite{Hedgecock75} of the rearing at 15$^\circ$C. Thus one needs separate dynamical variables---both the avoidance and the habituation---to model the temperature preferences at the rearing and at the assaying temperatures.  

Putting all of this together, we model the thermotactic bias $\Theta(t)$ as a a nonlinear function (here we choose $\tanh$) of the sum of the habituation and the avoidance in the droplet, $h(t)$ and $a(t)$. In turn, these integrate the experienced temperature (independent of and without food, respectively), and relax to zero (no preference or avoidance) with time if no additional reinforcement is present. (Note that this is distinct from the RW model, Eq.~(\ref{eq:rl}), where the association strength decreases only when a CS is observed, but not reinforced.) Similarly, by $h_{\rm r}(t)$ and $a_{\rm r}(t)$, we represent the habituation and the avoidance of the temperature, at which the worm was reared. The generalization is modeled by letting  the rearing and the in-droplet variables to become similar to each other with time. This results in a new, four-variable, generalization of the RW model, Eq.~(\ref{eq:rl}) (see \nameref{Section:model} for more details). While many similar model can be written, we only found one that could quantitatively fit the experimental data:
\begin{align}
 \tau_{h} \partial_t h&= A_h \Theta(t)-h + g_h h_{\rm{r}},\label{eqp}\\
  \tau_{a} \partial_t a&=(1-F(t))A_a \Theta(t)-a + g_a a_{\rm{r}},\label{eqn}\\
\tau_{h,\rm{r}} \partial_t h_{\rm{r}} &= -h_{\rm{r}},\label{eqpr}\\
\tau_{a,\rm{r}} \partial_t a_{\rm{r}} &= -a_{\rm{r}},\label{eqnr}\\
 \hat\Theta(t) &= \Theta_0 \tanh(h(t)-a(t)+c).
 \label{ttx}
\end{align}
Here $\tau_{h/a}$ are the time scales of the relaxation of the in-droplet habituation and avoidance, and $\tau_{h, {\rm r}/a, {\rm r}}$ denote these time scales at the rearing temperature. $A_{h/a}$ are the strength of conditioning of the corresponding internal states. $F(t)$ indicates the presence/absence of food at time $t$; it is always zero in the droplet by design. $\hat\Theta(t)$ is the thermotactic index predicted by the model, which is also the proxy for the warm/cold temperature that the worms experience.  $g_{h/a}$ are the strengths, with which the preferences to the rearing temperature generalize to the preferences in the droplet. We set the scale  of $g_{h/a}$ by rescaling $h_{\rm r}$ and $a_{\rm n}$, respectively. Then  $g_{h/a}=+ 1$ (or $-1$) for the warm (cold) worms. Finally, $\Theta_0<1$ is the maximum thermotactic index that the worm can experience due to the droplet geometry, and $c$ is the intrinsic thermal bias. See \nameref{Section:model} for a more detailed description.

Now we can develop intuition for how this model can account for the thermotactic dynamics data. For example, the worm reared at 15$^\circ$C initially is habituated to cold and hence starts with $h_r<0$, which generalizes to $h<0$; it is well fed, so that $a=a_r=0$. The worm then starts with $\Theta<0$. However, since the droplet contains no food, avoidance of the cold accumulates, $a<0$. At some point $a\approx h$, and $\Theta\approx0$. The worm then experiences both sides of the droplet nearly equally, and no biased reinforcement happens. However, if now $\tau_h<\tau_a$, the preference to the cold degrades faster than its avoidance, and $\Theta$ crosses zero and becomes positive. In contrast, if the assay starts following a short starvation period, then $a_r\neq 0$, which  generalizes to $a\neq 0$, and the worm's preference to the cold side starts from a smaller value. 

\subsection*{Fitting the model to data}
\label{Section:fitting_n2}

The following parameters of the constructed model must be fitted: four time scales, two reinforcement strengths, and initial conditions at the start of the droplet assay for every preference at the reared and the assay temperatures.  We fit the same parameters {\em globally} for all starvation durations at the same rearing temperature. We expect most parameters to be different for different rearing temperature and different mutants since worms develop into substantially different animals under these conditions  \cite{Xiao_2013, Goodman_Sengupta_2019}. However, not all of the parameter values are possible. For example, we expect that the bias decays for long times, and that the initial conditions, $a(t=0)$, depend monotonically on the starvation duration. We enforce such constraints, cf.~\nameref{Section:constraints}), into the optimization, cf.~\nameref{Section:optimization} to fit the model to the data.

Empirically, while many parameter values are not determined tightly by the data (i.e. have small effects on the quality of fit), some combinations of parameters are more strongly constrained. We use this observation and Bayesian model selection \cite{MacKay:2003wc} to reduce the number of model parameters, and hence to increase the generalization and decrease the optimization cost, cf.~\nameref{Section:reduction}. Most notable reductions include the same initial conditions for avoidance values at similar starvation durations, zero value of the thermotactic bias $c$, and the same ratio of habituation to avoidance time scales at the rearing and the assay temperatures, see \nameref{Section:reduction} and Tbls.~\ref{tab:N2-15},~\ref{tab:N2-25}. Figure~\ref{fig:wildtype} shows that this results in a model with excellent quantitative fits, with $\chi^2/f$ of 1.17 and 1.12 including all starvation conditions, for the cold and the warm worms, respectively. The fitted model confidence intervals, cf.~\nameref{Section:error_bars}, fall within the confidence intervals of the data, cf.~\nameref{Section:data}. Crucially, the model also reproduces salient qualitative features of the data, including rapid initial saturation, slower crossover, and---for 25$^\circ$C---the second reversal to zero bias. Further, the decay and the eventual reversal of the initial $\Theta$ as a function of the starvation duration is also modeled well, Fig.~\ref{fig:wildtype}. While most inferred model parameters, Tbls.~\ref{tab:N2-15} and \ref{tab:N2-25} are not illustrative, a few parameter combinations are notable. First, the ratio $\tau_h/\tau_a$ is reliably small, $\approx 0.28$ and $\approx 0.41$ for the cold and the warm worm, respectively. That the habituation has faster dynamics than the avoidance is what allows for overshoots and oscillations within our model, and should guide future experiments. Second, the bias $c$ is zero for the cold worm and is negative for the warm worm. This is consistent with the worms avoiding warmer, but not colder temperatures in isothermal tracking  \cite{Yamada03}, and is an independent confirmation of inferences from our model. 

\subsection*{Thermotactic dynamics in mutants}
To test the generality of our model, we investigated worms with mutations in the insulin-like signaling pathway, which is involved in olfactory, chemotactic, and thermotactic avoidance learning \cite{Chen2012,Tomioka2006,Kodama06}. We chose three mutants that affect insulin-like signaling at different stages, see \nameref{Section:strains}. These show qualitative differences in their thermotactic dynamics compared to the wild-type worm. As discussed earlier, {\em ins-1(nr2091)} worms learn a thermal preference, and this preference persists for the duration of the experiment (Fig.~\ref{fig:mutants}A). {\em daf-2(e1370)} have a mutation in an insulin-like receptor which produces a temperature-sensitive dauer phenotype \cite{Gems_1998}, disrupting the pathway predominantly at warm temperatures (Fig.~\ref{fig:mutants}B). Consistent with this, warm {\em daf-2} worms show a persistent thermal preference like {\em ins-1} worms, while cold worms have a reduced avoidance response, but with a similar timing. {\em age-1(hx546)} mutants, with a degraded function of a kinase downstream of the insulin-like peptide receptor, have accelerated rates of negative association in their thermal learning assay \cite{Kodama06}. Indeed, these mutants show faster crossover for both rearing temperatures (Fig.~\ref{fig:mutants}C). 

If our model captures biological mechanisms, we expect the mutants to have have different parameters values in its avoidance branch, Eqs.~(\ref{eqn}, \ref{eqnr}), but not the habituation branch, Eqs.~(\ref{eqp}, \ref{eqpr}). Therefore, we fit the mutant data by keeping all but the avoidance parameters to be the exactly those  inferred for the N2 worm, cf.~\nameref{Section:fitting_n2}. We only fit the avoidance parameters for each mutant and each rearing temperature, see \nameref{Section:mut_const}, \nameref{Section:Mutant_fits}. Further, we do not fit {\em any} parameters for the {\em ins-1} worm, and simply remove the avoidance branch for this mutant, setting $a(t)=a_r(t)=0$, consistent with the avoidance effectively removed by the {\em ins-1} mutation. The agreement between the model and the data for all mutants in Fig.~\ref{fig:mutants} is generally within the error bars, cf.~\nameref{Section:error_bars}. This is better than could have been expected, considering that mutant worms are very different from the wildtype and implies a biological relevance rather than purely statistical prowess of the model. The inferred values of the parameters, Tbl.~\ref{tab:mut-avoidance}, are consistent with the expected effects of the mutations. Crucially, the ratio $\tau_a/\tau_h$ remains large for all datasets. 

To further explore the biological relevance, we fit an alternative model where only the habituation (rather than the avoidance) parameters are fitted for mutants,  contradicting the known biology.  Crucially, this alternative model, Tbl.~\ref{tab:mut-habituation}, fits the data {\em quantitatively} worse than the primary model, cf.~Tbl.~\ref{tab:mut-fitvals}, so that {\em a posteriori} odds are about $10^{16}\!:\!1$ in favor of the primary model. Such agreement between the known biology and the quantitative analysis again signals that the primary model captures the relevant biological mechanisms. 

\section*{Discussion}
We combined novel experimental tools for fast behavioral assaying, mutants for genetic perturbations, and mathematical modeling to understand the thermal preference dynamics of {\em C.~elegans}. We discovered that (i) the dynamics is governed by two independent pathways: positive association (habituation) to the experienced temperature independent of food, and negative association (avoidance) of the temperature with no food; that (ii) the avoidance time scale is slower than the habituation one; that (iii) the dynamics is quantitatively different in warm and in cold worms; that (iv) the full dynamical description requires at least four dynamical variables; and that (v) these dynamical variables are likely related to the worm's ability to generalize across similar temperatures. We {\em quantitatively} fit all of the available data (different life histories and genetic background) with high precision, indicating that the model has captured the underlying biological mechanisms.

Small differences in experimental conditions (assay duration, gradient shape, range, and steepness, rearing protocols, etc.) result in drastic changes in {\em C. elegans} thermotaxis  \cite{Hedgecock75,Yamada03,Kimata2012,Jurado2010,Luo2014}. This makes it hard to definitively address open questions, such as the effect of food on the thermotactic response \cite{Chi07,Yamada03,Luo06}. While careful experiments that measure the life history of temperature and food exposure in individual worms will be needed to complete the details, the existence of multiple reinforcement pathways, with the fastest of them operating on scales of less than half an hour, is consistent with many of these conflicting observations. Our discoveries support the existence of an active avoidance of thermal memory when food is absent \cite{Hedgecock75}. They also demonstrate that food is not necessary for the establishment of the temperature preference consistent with other studies \cite{Chi07,Yamada03,Luo06,Biron06,Hedgecock75,Ryu02,Clark06}. Further, our multidimensional framework can explain the previously ignored non-monotonic dependence of the temperature preference on the exposure duration \cite{Biron06,ramot_thermotaxis_2008}. Finally, it may account for the limited operating range of thermotaxis \cite{ramot_thermotaxis_2008} as weaker generalization over larger temperature ranges.

Our model, Eqs.~(\ref{eqp}-\ref{ttx}), is inspired by the animal learning literature \cite{RescorlaWagner,Niv:2009dw}, and may seem foreign in the field of {\em C.~elegans} thermal memory. We suggest that the aforementioned difficulties in explaining diverse experimental data come from the focus of traditional models on {\em one} dynamical preferred temperature, while our data demonstrate that there are, at least, two separate conditioning pathways. The traditional ``one preference'' model can be augmented to account for this. For example, the animal may be modeled to have a food-independent preferred temperature and a food-dependent temperature to avoid, with the joint push-pull of the two establishing the observed thermotactic pattern. Another possibility is that the animal may have a set temperature, but also an internal state (or states) describing the strength and the nature (attraction or repulsion) of the temperature drive. The simplest versions of either of these models would have too few dynamical variables to explain our data, cf.~\nameref{Section:dimensionality}. To attain the necessary complexity, we would need to consider multi-scale temperature averaging,  or more than two preference/avoidance temperatures and complicated generalization rules. With little known biology to build on, such models would be indistinguishable from our functional, Rescorla-Wagner style model. Thus next crucial experiments should move away from the ``one preference'' paradigm and focus on characterizing the dimensionality and the nature of thermal preference states, and the ability to generalize among them. Such experiments will likely benefit from modern interpretable machine learning techniques \cite{Daniels:2015ht,daniels_pnas}, which can automate the search for a correct multi-dimensional model.

Two distinct mechanisms for positive and negative associations are biologically plausible (activation and excitation generally are mediated by different neurons and neuropeptides), but infrequently explored across the animal kingdom. In particular, while dopamine is implicated in associative learning in vertebrates \cite{Schultz98}, very little is known about what can mediate negative associations \cite{Schultz2000}. Similarly, in modeling, avoidance is often represented as a negative attractive signal. However, it is precisely the separation of the two pathways that allows the worm to exhibit oscillatory thermal preference, including spontaneous recovery of previously extinguished associations, a hallmark of conditioning  \cite{Pavlov-27}. It is also the major contributor to our model's ability to simultaneously explain spontaneous recovery (e.g., Fig.~\ref{fig:mutants}C), asymmetric responses to appetitive vs.\ aversive cues and to conditioned association vs.\ conditioned inhibition, and, finally, generalization among similar, but not equivalent cues.  Recent work has identified genetic \cite{Mohri_2005, Goodman_Sengupta_2019} and neurophysiological \cite{Mori95, Kimura04, Hawk_2018} factors underlying {\em C.~elegans} thermotaxis and thermal preference. We thus hope that our discoveries in the worm, and especially identification of the two independent learning mechanisms, will spur a search for similar phenomenology in larger animals.

Finally, it is intriguing to speculate about the functional importance of two independent thermal preference pathways, with the avoidance being reliably slower. When searching for food, animals start where they expect food to be found based on the conditioned stimuli. However, if the search there is unsuccessful, they turn to searching {\em elsewhere} (avoiding the already searched area) rather than {\em everywhere} (including the area just searched). Switching to {\em elsewhere} involves complex statistical strategies, such as Levy flights \cite{bartumeus}, which the worm exhibits as well \cite{salvador}. A system with a single association strength would result in no preference and thus in searching {\em everywhere} as the association decays. In contrast, a distinct avoidance channel with a long time scale would guide the animal away from the condition with no food. Thus we expect that the multi-modal and multi-scale thermal preference is optimal for fast search in temporally varying environments. 

\section*{Materials and Methods}
\subsection*{Nonlinear dynamics of animal learning}
\label{Section:learning}

Here our goals are two-fold. First, we would like to show that traditional animal learning models map into the dynamics similar to Eq.~(\ref{eq:rl}). Second, we want to show that such dynamics, with realistic mapping between measurable variables and variables internal to the animal cannot produce oscillations in our experimental systems (and, in particular, zero crossings that we see in Fig.~\ref{fig:wildtype}). 

We start with the Rescorla-Wagner model of conditional associations \cite{RescorlaWagner,Niv:2009dw}:
\begin{equation}
    V_i(t+ \Delta t)= V_i(t) + \eta \delta_i(t) \left(\lambda_0\delta_\lambda(t) - \sum_j V_j(t)\delta_j(t)\right),
\label{eq:RW}
\end{equation}
where $V_i(t)$ is the strength of association between the $i$th Conditioned Stimulus (CS) and the Unconditioned Stimulus (US),  $\eta$ is the learning rate,  $\lambda_0$ the magnitude or salience of a single US reward, and $\Delta t$ is the duration of one experimental epoch. Further, $\delta_i(t)=1$  if the $i$th CS is present at time $t$, and 0 otherwise. Similarly, $\delta_\lambda(t)=\{1,0\}$ depending on if the US was or was not present at time $t$.  In other words, $\sum_j V_j(t)\delta_j(t)$ is the US prediction based on the strength of all CSs present at the time. Thus if a CS is present, then Eq.~(\ref{eq:RW}) posits that its US association strength is changed in proportion to how well all of the CSs predict the realized US reward, $\lambda_0\delta_\lambda(t)$. 

We define the frequency of the CS and US presentations, $f_i=\langle \delta_i\rangle$ and $f_\lambda=\langle \delta_\lambda\rangle$, as well as the frequency of US given the $i$th CS $f_{\lambda|i}=\langle \delta_i\delta_\lambda\rangle/f_i$ and one CS given another $f_{j|i}=\langle \delta_i\delta_j\rangle/f_i$. We then introduce a time interval $dt$, which is small enough so that the association strength does not change much over its duration, $\eta\lambda_0f_i f_{\lambda|i} \frac{dt}{\Delta t}\ll 1$, and which nonetheless is large enough so that it contains a lot of CS and US presentations, $f_if_{\lambda|i}\frac{dt}{\Delta t}\gg 1\langle \delta_i\rangle \langle \delta_\lambda\rangle\Delta t\gg 1$. We integrate Eq.~(\ref{eq:RW}) over this interval to obtain
\begin{align}
  \sum_{t'\in(t,t+dt)}  \left(V_i(t'+\Delta t)-V_i(t')\right)&=  dt \frac{\eta/\Delta t}{dt/\Delta t} \sum_{t'\in (t,t+dt)}  \left(\lambda_0 \delta_\lambda(t')\delta_i(t')- \sum_j V_j(t') \delta_i(t')\delta_j(t')\right),\\
  dV_i(t) &= dt \frac{\eta}{\Delta t} f_i \left(\lambda_0 f_{\lambda|i}(t) - \sum_j V_j(t)f_{j|i}(t) \right), \\
  \tau \frac{dV_i}{dt} &=  f_i(t) \left(\lambda_0 f_{\lambda|i} - \sum_j V_j(t)f_{j|i}(t) \right),
    \label{eq:RL_deriv}
\end{align}
where $\tau = \Delta t/\eta$ is the learning time scale. 

Sometimes all CSs are exclusive: for example, while many temperatures can serve as predictors of food, an animal can only experience one temperature at a time. In this case, $f_{j|i}=0$ for $j\neq i$, and $f_{i|i}=1$. Then Eq.~(\ref{eq:RL_diff}) becomes
\begin{equation}
  \tau \frac{dV_i}{dt}=  f_i(t)\left(\lambda_0f_{\lambda|i}(t)-V_i(t)\right).
    \label{eq:RL_diff}
\end{equation}
This is  Eq.~(\ref{eq:rl}) in the main text. Note that, for Pavlovian associations, $f_i(t)$ is under the control of an experimenters, while the animal can influence it in the operant conditioning protocols, so that  $f_i(t)=f_i(V(t))$. Also note that in traditional analysis of conditioning, the CS must precede (and hence predict) the US for the association to form. In the context of our experiments, this temporal contingency structure can be disregarded because the CS temperature signals are experienced by the worm not episodically, but constitutively. 

We now turn to showing that, in the case of our experiments, this version of the Rescorla-Wagner model cannot produce the experimentally observed crossings of the zero thermal bias line. First, notice that, because of the finite temperature resolution by the worm (cf.~\nameref{Section:model}), there are only about two discernible CSs in the droplet: warm ($+$) and cold ($-$). Further, when the worm is not in the warm, it is in the cold, so that $f_+=1-f_-$. In its turn, the worm's position and hence the experienced CS is affected by the strength of the associations, so that $f_+=f_+(V_+-V_-)$.  In the droplet, there is no US (no food), so that $f_{\lambda|\pm}=0$. Thus in the droplet, the model in Eq.~(\ref{eq:RL_diff}) becomes:
\begin{align}
    \tau \frac{dV_+}{dt}&= -f_+(V_+-V_-)V_+,    \label{eq:two_sides1}\\
    \tau \frac{dV_-}{dt}&= -(1-f_+(V_+-V_-))V_-.
    \label{eq:two_sides2}
\end{align}
We now note that $f_+\approx 0.5$ corresponds to a small thermal bias $f_+-f_-=1-2f_+\approx 0$. The small bias is only possible when $V_+\approx V_-$. In this regime, Eqs.~(\ref{eq:two_sides1}, \ref{eq:two_sides2}) further simplify:  
\begin{equation}
    \tau \frac{dV_\pm}{dt}= -0.5V_\pm.\label{eq:simplifiedRW}
\end{equation}
These are equations for a simple exponential relaxation $V_\pm\to0$, and both association strengths will decay with the same rate $0.5/\tau$. Thus if the thermal bias is near zero at some point in time, it must remain so indefinitely, and substantial oscillations are not possible. 

One can create a single crossing of the zero thermal bias line using Eq.~(\ref{eq:simplifiedRW}) if the time scales of the dynamics of the two associations are not the same, $\tau_+\neq \tau_-$. However, this is likely insufficient to our experiment for three reasons. First, some of the mutant worms exhibit more than one zero crossing, cf.~Fig.~\ref{fig:mutants}. Second, the cold and the warm sides of the droplet are less than a degree apart, which is much smaller than the range of temperatures the worm tolerates. Hence one would expect differences in time scales, if present, to be similarly relatively small. In this case, the overshoots of the zero line would be tiny and would take a long time to develop, while in all of our experiments the overshoots happen on the same time scales of the overall dynamics, cf.~Figs.~\ref{fig:wildtype} and \ref{fig:mutants}. Third, the worm is only reared at one temperature. Thus in the simplest model, we would have either $V_+=0$ or $V_-=0$ at zero time. Thus a small temperature bias would require not just $V_+=V_-$, but also $V_+=V_-=0$.  In this case, even two distinct dynamical scales would not produce zero crossings. Collectively, these arguments suggest that a simple Rescorla-Wagner style model cannot account for the worm's thermal preference dynamics, and more complicated models are needed.

\subsection*{Strains and preparation}
\label{Section:strains}
 The mutant strains used in this study were as follows: We obtained strains (N2, {\em ins-1(nr2091)}, {\em daf-2(e1370)}, and {\em age-1(hx546)}) from the Caenorhabditis Genetics Center at the University of Minnesota.

All experiments used young adult animals cultivated at 15$^\circ$C and $25^\circ$C on nematode growth medium (NGM: 50 mM NaCl, 15 g/L agar, 20 g/L peptone, 1 g/L g/L, 1mM cholesterol, 1mM $\mathrm{CaCl}_2$, 1mM $\mathrm{KH_2PO_4}$ agar plates seeded with  {\em Escherichia coli} strain OP50 under standard conditions \cite{Brenner_1974}. M9 Buffer (3g $\mathrm{KH_2PO_4}$, 5 g NaCl, 6 g $\mathrm{Na_2HPO_4}$, 1 ml 1 M $\mathrm{MgSO_4}$, $\mathrm{H_2O}$) used for strain washes and assay. Animals were stage-synchronized using a standard bleach synchronization protocol \cite{Stiernagle_06}. We washed synchronized young adult animals with 1ml of M9 buffer into a 15mL Falcon tube, added an additional 10mL of M9 buffer, and pelleted animals by spinning at 0.4 RCF for 1 minute. We aspirated the supernatant and repeated the wash. In the case of ‘well-fed’ state experiments, we re-suspended the animals in 2ml of M9 and poured them onto a 5cm NGM plate. We then picked individuals into 2ml of buffer on a second 5cm NGM plate and transferred them into the $\mu$Droplet assay for observation. In the case of ‘starved state’ experiments, we decanted animals onto a 10cm NGM plate and allowed them to starve at their rearing temperatures as per protocol durations. We then washed animals off the plate using 2ml of M9 and decanted them onto a 5cm NGM plate. We picked individuals into 2ml of buffer on a second 5cm NGM and the transferred them into the $\mu$Droplet assay for observation.

\subsection*{$\mu$ droplet assay}
\label{Section:droplets}
Individual worms are picked into a grid of 3$\mu$L M9 buffer droplets on glass-printed 4mm hydrophilic spots (Electron Microscopy Sciences Inc.\ Item $\#$63430-04) surrounded by a hydrophobic PTFE (Teflon) surface. A coverslip placed on top of a 127 $\mu$m-thick silicone gasket bonded to the assay slide with VALAP (1:1:1 Vaseline, Lanolin, Paraffin).

The assay is centered and clamped on a temperature controlled aluminium stage. The stage is preset to either $15^\circ$C or $25^\circ$C depending on rearing temperature. Animals are acclimated for 5 minutes before imaging commences. The black-anodized Aluminium stage measures 165x58x3mm. Waterblocks (Swiftech MCW30) are secured to the ends of the stage through 40x40x3mm aluminium spacers coated with thermal paste. A 40x40mm peltier element (MCTE1-19908L-S) is secured under one spacer and waterblock. A waterbath circulator (Fisher Scientific IsoTemp 3016) is used to control the initial stage temperature. The peltier element is used to establish and program the temperature gradient. A 15-$25^\circ$C stage temperature translates to a $1^\circ$C/cm assay steepness and thus a temperature range of 19.6 to $20.4^\circ$C on the gradient-aligned extremes of the droplet confirmed through direct IR camera observation and COMSOL Multiphysics modeling (data not shown).

Illumination for image capture is provided by two red LED light strips (3W 48-LED 180-Lumen Aluminum alloy light strip) positioned $\sim9$cm above the  imaging stage. Raw monochrome images are captured via a DSLR camera (Nikon D7000) with a fixed macro lens (Nikon AF Micro Nikkor 60mm f/2.8D) controlled by Nikon’s Camera Control Pro software. In order to minimize the correlation of the worms’ position through time, capture rate is 6 frames per minute during long-term observation.

Image capture begins at the 5-minute mark in the absence of a temperature gradient for 100 frames ($\sim 16$ minutes), after which a 15-$25^\circ$C stage gradient is applied and recording continues to 1440 frames (4 hours). The absence of a gradient for the first 100 frames was used as a built-in control for each assay trial.

Experiments were performed successively 2-3 times per day. No randomization procedures were undertaken to control for time-of-day effects.

\subsection*{Data processing}
\label{Section:data}
We post-processed images and computed behavioral metrics using MATLAB. For this, we located regions of interest (ROI) around each droplet, subtracted the background, and produced a binary image of the worm. We used the measure of worm area to filter out potential segmentation artifacts before calculating the center of mass position of the worm for each frame. The ROI determines the extremes of possible motion and is used to normalize the animal movement along a linear index from -1 ($19.6^\circ$C) to +1 ($20.4^\circ$C). When artifacts occur, such as in the identification of two objects in a droplet, we do not record a position in that frame. If a worm track is less than 95\% complete, we discard it from the data set. A final thermal preference metric, the thermotaxis index, $\Theta$, is calculated by summing all normalized values of an animal’s movement in a certain time and dividing by the total number of observations.

Specifically, for each worm type (rearing temperature, mutation), all data are indexed by three indices:  $\mu$ stands for the condition (i.~e., the starvation duration), $n$ stands for the $n$th individual worm in that condition, and  $t$ represents the time in hours since the beginning of assaying in the droplet. Since occasionally a worm is not tracked for some times due to image processing artifacts, the number $N_\mu(t)$ of individual worms tracked at time $t$ in condition $\mu$ is time dependent.

The trajectory on the gradient of the $n^{th}$ individual worm in condition  $\mu$  at time $t$ is $\Theta_{\mu,n}(t)$. Therefore, the average thermotactic index at time $t$ is 
\begin{equation}\bar{\Theta}_\mu(t) = \sum_{n=1}^{N_\mu(t)}\frac{\Theta_{\mu,n}(t)}{N_\mu(t)}.
 \end{equation}
 Similarly, the variance of thermotactic index at time $t$ is
 \begin{equation}{\rm var}_\mu(t) = \sum_{n=1}^{N_\mu(t)}\frac{(\Theta_{\mu,n}(t)-\bar{\Theta}_\mu(t))^2}{N_\mu(t)-1}.
\end{equation}
Then the standard error of the mean for worms in condition $a$ at time $t$,
\begin{equation}s_\mu(t) = \left(\frac{{\rm var}_\mu(t)}{N_\mu(t)} \right)^{1/2},
\end{equation} 
defines the experimental error bars on the thermotactic trajectory for the time and the condition.

As a measure of the overall noise of the data for worms in condition $\mu$, we use the time averaged variance of the data trajectories over all worms in the  condition $\mu$
\begin{equation} \bar{v}_\mu = \frac{1}{T} \sum_{t=0}^{T} v_\mu(t),
\end{equation}
where $T=4$ hours is the assay duration in the droplet.

Finally, for presentation purposes only (but not for the fitting), to remove rapid fluctuations, in all figures the thermotactic index and its experimental error are filtered through a causal exponential filter with the time scale of 6 min. 

\subsection*{Bounding the dimensionality of the thermal memory dynamics}
\label{Section:dimensionality}
{\em C.~elegans} thermal memory is a dynamical system. To model the behavior, we first must estimate the number of dynamical variables in this system. Suppose that the dynamics, averaged over all worms, can be fully described by a $d$-dimensional time dependent vector $\vec{x}(t)$ that evolves according to an unknown, but continuous and deterministic dynamics. Suppose further that the one-dimensional observable $\Theta(t)$, the thermotactic index indicating the current temperature the worm guides itself to, is a smooth function of $\vec{x}(t)$, which may depend on all components of $\vec{x}(t)$. We estimate the dimensionality of $\vec{x}(t)$ using the following arguments.

\paragraph*{Thermotactic memory dynamics is multidimensional}
Suppose that $d=1$. Then, $x(t)$ and $\Theta(t)$ are functions of each other, and, for the dynamics to be well defined, the velocity $\Theta'(t)$ should be a single valued function of the position of the worm $\Theta(t)$. This is not the case. For example, at $\Theta(t)=\mp 0.1$, the thermotactic index $\Theta(t)$ for well-fed wild type worms at $15,25^\circ$C in Fig.~\ref{fig:wildtype} has two different thermotactic velocities $\Theta'(t)$ early and late in the experiment. Even simpler, the thermotactic index crosses a zero multiple times. Thus $d=\dim(\vec{x}(t)) > 1$.

\paragraph*{Thermotactic memory dynamics has $d\geq2$}
To determine if $d=2$ would be sufficient to capture the behavior, we consider the delayed embedding representation of the dynamics \cite{Takens81} as a $k$ dimensional vector  $\vec{\Theta}_k(t)= (\Theta(t), \Theta(t-\tau), \cdots, \Theta(t-(k-1)\tau)$. The Takens theorem \cite{Takens81} allows reconstruction of a $d$ dimensional attractor from such $k=2d+1$ dimensional embedding. However, if $\tau$ is much smaller than the characteristic time scale of the dynamics, then the $d$ dimensional dynamics can be reconstructed uniquely from even smaller sequences, $\vec{\Theta}_k$ with $k=d$, provided the trajectories in the  $\vec{\Theta}_k$ space do not self-intersect (to see this, notice that, at $\tau\to0$, $\vec{\Theta}_k$ maps one-to-one onto $(\Theta(t), \Theta'(t), \Theta''(t),\dots,\Theta^{(k-1)}(t)$). Thus to bound the dimensionality of $\vec{x}$, we seek the minimal $k$, for which delayed embedding trajectories at small $\tau$ show no self-intersections. The characteristic time scale of the thermal memory dynamics is $\mathcal{O}(1 {\rm h})$ (cf.~Fig. \ref{fig:wildtype}), and the trajectories show meaningful changes, statistically distinguishable from noise, for $\tau>\mathcal{O}(1 {\rm s})$. This allows for a broad range of $\tau$ for our analysis. In what follows, we choose $\tau=6.67$min (every 40th data point), but results are qualitatively the same for similar values. To quantify the uncertainty in the delayed embedding trajectories due to the experimental noise, we generate bundles of $n_{\rm t}=20$ trajectories that are different by their starting time, $t_0 = 15 {\rm min} + \frac{\tau}{2n_{\rm t}}i$, $i=0,\dots,n_{\rm t}-1$, and we look for self-intersections of these trajectory bundles. 

\begin{figure}[h!]
\includegraphics[width=10cm]{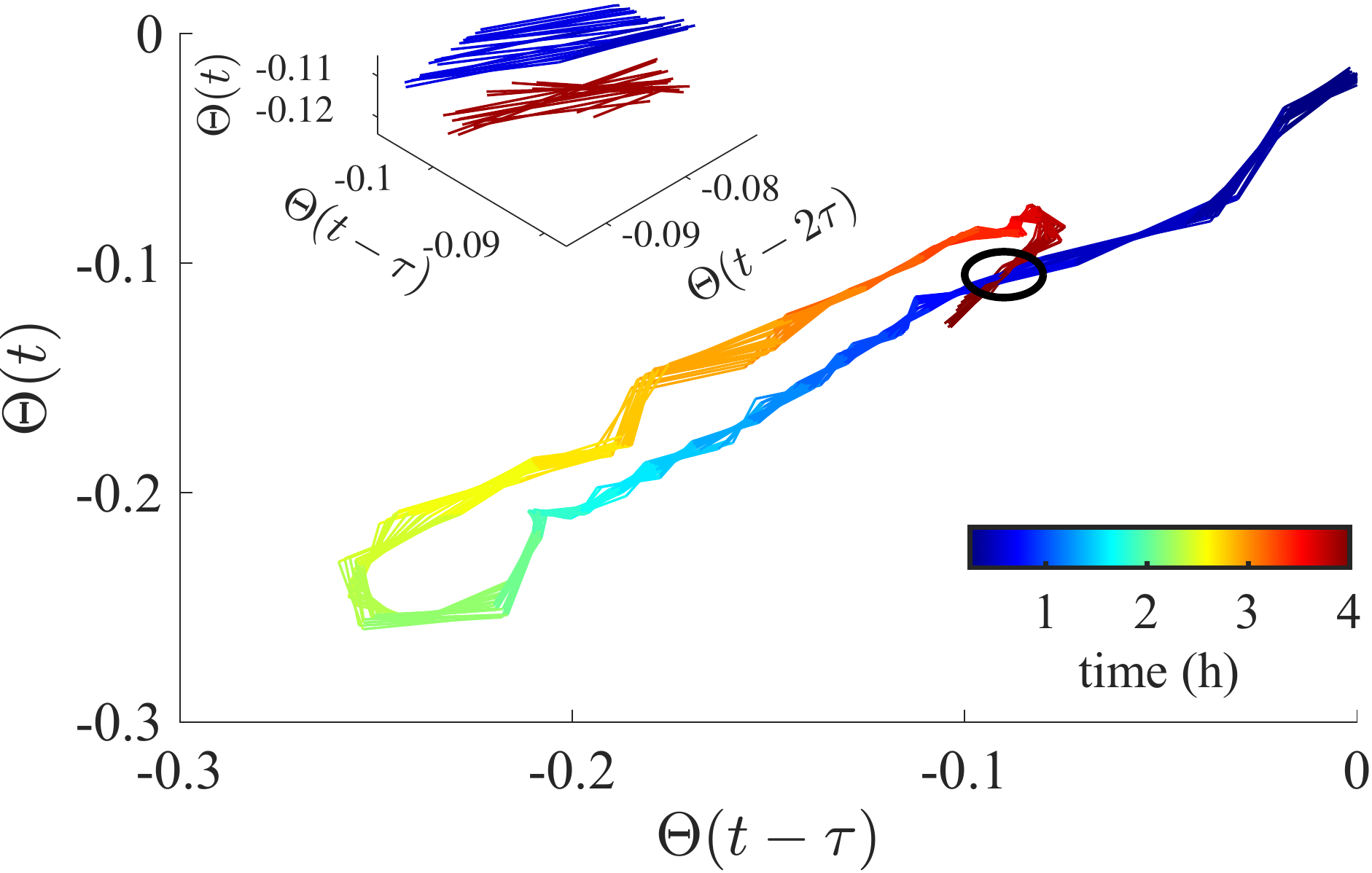}
\caption{\textbf{Delayed embedding analysis of the thermotactic preference dynamics.} Thermotactic index of the mean wild type N2 worm reared at $25^\circ$C and starved for half an hour before assaying the thermotactic response in the droplet is analyzed using the delayed embedding coordinates. The time within  the experiment is indicated by the color bar starting from 15 minutes (dark blue) and ending at 4 hours (dark red). The delay time $\tau\approx 6.67$  min for both the two and three dimensional embedding. Bundles of curves contain 20 curves each, corresponding to different starting offsets. The two dimensional embedding (main plot) has a self-intersection, indicated by a black circle. This confirms that the underlying dynamics is, at least, three dimensional. Zooming in on the relevant region in a three dimensional embedding (inset), we  find no self-intersection.}
\label{fig:Taken}  
\end{figure}

Figure \ref{fig:Taken} shows the trajectory bundles with $k=2,3$ for the 25$^\circ$C worms starved for 1 hour. The 2-d embedding shows a clear intersection of the bundles, while the 3-d embedding does not. Performing the same analysis for all other starvation durations and both rearing temperatures (15$^\circ$C and 25$^\circ$C), we observe that $k=3$ is always sufficient to avoid self-intersections of the bundles. Thus we conclude that $d\geq 3$.

\paragraph*{Linear analysis suggests $d\geq 4$}
For the well-fed and the 30 minutes starved worms raised at $25^{\circ} $C, the observed thermotactic index $\Theta(t)$ is always far from the maximum value of the index,  $\Theta_0\approx 0.27$, observed for  $15^{\circ} $C (cf.~Fig \ref{fig:wildtype} B). We thus treat these trajectories as far from saturated, and assume that they can be produced by the dynamics linearized around $\Theta=0$. These trajectories have different oscillation periods: about 3 h for the well-fed worm and ${>4}$ h for the starved ones. For a linear dynamical model, eigenvalues describing the oscillations come in complex conjugate pairs, each pair sharing the frequency. Thus two complex conjugate pairs are required to model the two distinct oscillation frequencies. We thus conclude that $d\geq4$, which is consistent with all of the previous analyses.

\subsection*{Model}
\label{Section:model}
The same dynamical system can be represented in many different ways. Our focus is on finding a description that is interpretable, biologically plausible, and can be probed through realistic experiments. We aim for a parsimonious description, and hence will look for a model with $d=4$ dynamical variables, the smallest number of variables consistent with the analysis above.

The identity of the relevant internal states of the worm that store its thermal preference remains unclear \cite{Chi07}. While the only measurable quantity in our experiments is the thermotactic index, it is not an internal state. In diverse experiments, the worm develops either a preference to or avoidance of certain temperatures \cite{Hedgecock75}, and there has been a lively debate in the literature \cite{Chi07,Yamada03,Luo06,Biron06,Hedgecock75,Ryu02,Clark06} regarding whether the avoidance or the preference is primary, and if either is conditioned on the presence of food or not. Knowing from the analysis above that the thermal preference dynamics is, in fact, multidimensional allows us to avoid the controversy, and to try various forms of the internal states. We explored many such models, and the only model that we found, which was able to fit the entire corpus of our data {\em quantitatively}, has separate dynamical variables representing which temperature the worm likes, and which temperature it avoids. Crucially, in this model, the attractive temperatures are not conditioned by food, but describe habituation: the worm likes the temperature that it has experienced before, which can be viewed as conditioning on prior survival. In contrast, the avoidance is conditioned by the {\em absence} of food: the worm does not like the temperature where the food has been unavailable previously. We stress that this is different from most models in the literature, which usually have only one dynamical variable, representing preference conditioned on the presence of food \cite{Biron06}.

In principle, within the conditioned associations paradigm, one may model the habituation and the avoidance as functions of the temperature field, so that every temperature has its own habituation and avoidance values, and the {\em generalization} of preferences to the nearby temperature ensures that these functions are smooth. However, the $\mu$ droplet assay establishes the gradient of only $0.8^\circ$C across the entire droplet. The worms move stochastically, and their length is comparable to the droplet diameter. Thus geometrically they cannot stay at the poles of the droplet, where $\Theta=\pm1$. Because of this, $\Theta$ usually spans only about a quarter of the possible range, between about $\pm 0.25\dots \pm -0.3$. Thus the worms typically experiences only $\sim0.2^\circ$C temperature range, which is comparable to their thermal sensitivity of $0.1^\circ$ \cite{Mori95,Kimura04,Clark06}. In other words, the worms cannot reliably distinguish more than about two temperature values in the droplet: warm and cold. In this situation, modeling the habituation and the avoidance as functions of the temperature field is unnecessarily complex, and we model the worm's internal states with two scalar variables. First, $h$ (unconditioned positive preference, or habituation) takes positive values when the worm likes the warm side. Second, $a$ (conditioned negative preference, or avoidance) is positive when the worm avoids the warm side. We then define the overall temperature preference as the combination of the habituation and the avoidance, passed through a saturating nonlinearity, as in Eq.~(\ref{ttx}). 

With this, we identify our four dynamical variables with habituation and avoidance at the rearing temperature and at the assay temperature.  The rearing variables change autonomously during the droplet assay, while the assay temperature variables are affected by the animal's behavior.  The dynamics is given by Eqs.~(\ref{eqp}-\ref{ttx}).

Note that the rearing temperature preferences decay autonomously in Eqs.~(\ref{eqpr}, \ref{eqnr})  since these temperatures are not experienced in the assay. However, during the rearing time, the dynamics of $h_{r}$ and $a_{r}$ are expected to have the conditioning terms similar to $h$ and $a$ in Eqs.~(\ref{eqp},~ \ref{eqn}). Finally, for parsimony, we do not model the generalization from the droplet to the rearing temperature.

It is useful to explain the dynamics described by Eqs.~(\ref{eqp}-\ref{ttx}) narratively. A worm's preference of a certain temperature emerges from a combination of an unconditioned habituation and a conditioned avoidance, the latter driven by the absence of food. Worms habituate to their rearing temperature. If the period of rearing is followed by starvation before being placed in the droplet, then additionally the avoidance of the rearing temperature sets in, roughly in proportion to the starvation time. Experiences during rearing are fully under the control of the experimenters, and thus represent classical or Pavlovian conditioning \cite{Pavlov-27}. In contrast, the habituation and the avoidance of one side of the droplet are initially determined by generalizing the relevant preferences from the rearing temperature. However, as the worm begins to explore the droplet, it controls the temperature that it experiences, setting up an operant conditioning protocol, in which it habituates to the side of the droplet that it visits the most. At the same time, since there is no food in the droplet, the worm also develops the avoidance of the side it frequents. Crucially, this happens in the background of the decaying habituation and avoidance to the rearing temperature (and hence of the generalization effects), which cannot be reinforced while in the droplet and hence decay as solutions of the first order linear differential equation---exponentially with time.

\subsection*{Constraints on the model parameters}
\label{Section:constraints}

Here we discuss constraints on the model parameters, which we use in model fitting.

\paragraph*{Constraints on initial conditions} All worms in the experiment were raised for days, well-fed at their rearing temperature. In contrast, the typical time scales of the thermal preference is hours (cf.~Fig.~\ref{fig:wildtype}). This implies that the habituation to the rearing temperature is likely saturated for all worms. In other words, for all starvation conditions, $h_{\rm{r}}(0)$ is the same,  depending only on the rearing temperature and the worm genetic background. The dynamics of habituation in the droplet in Eq.~(\ref{eqp}) is driven by the generalization for the entire duration of the rearing, and is likely to be saturated as well, at some value $h(0)$, the same for all starvation conditions, but also likely different for different rearing temperature and the genetic background. Further, $h(0)$ is potentially different from $A_h$ (saturation value in the droplet) since the maximum strength of conditioning may depend on the ambient temperature.

In our model, avoidance is a non-decreasing function of the starvation duration $s$. A typical starvation duration is a few hours long, so the avoidance states are probably not saturated. Thus the magnitude of the initial avoidance of the rearing temperature for different $s$, $|a_{s,\rm{r}}(0)|$, for worms raised in the same temperature, should be a non-decreasing function of $s$, $|a_{s_1,\rm{r}}(0)|\le |a_{s_2,\rm{r}}(0)|$ for $s_1<s_2$. Since avoidance is caused by starvation, the initial avoidance for a well-fed worm is zero, $|a_{0,\rm{r}}(0)|=0$. The initial avoidance in the droplet emerges from the generalization of the avoidance at the rearing temperature, and thus should follow the same law: $|a_{s_1}(0)|\le |a_{s_2}(0)|$ for $s_1<s_2$, and $a_0(0)=0$. Further, since the definition of the thermotactic index in the droplet sets $\Theta(0)=0$ at $20^\circ$C, all initial conditions for worms raised at $15^\circ$C will be non-positive and for those trained at $25^\circ$C will be non-negative.

In summary, there are two initial conditions (rearing and droplet) for the habituation for each rearing temperature. There are 4 different starvation durations (and the well-fed worm) for the $15^\circ$C worm, which corresponds to 8 ordered initial conditions for rearing and droplet avoidance, and 3 different starvation durations (in addition to the well-fed) for the $25^\circ$C worm, resulting in 6 ordered avoidance initial conditions.  

\paragraph*{Constraints on dynamical parameters} For parsimony, the habituation and avoidance time scales $\tau_h$, $\tau_a$, $\tau_{h,\rm{r}}$, $\tau_{a,\rm{r}}$, the maximum conditioning strengths $A_h$ and $A_a$, and the bias $c$ are assumed to be common among all starvation durations reared at the same temperature. From Fig.~\ref{fig:wildtype}, we observe that the mean thermotactic index never goes above $\approx 0.27$, so we choose $\Theta_0=0.27$ for all starvation and rearing temperatures to simplify the multidimensional fitting. 

Since worms generally avoid high temperatures \cite{Yamada03}, the thermotactic index for starved worms at $25^{\circ}$C have a negative bias after $\sim3$ hours, cf.~Fig.~\ref{fig:wildtype}. Such bias is not present at  $15^{\circ}$C. Thus we choose to set $c=0$ for $15^{\circ}$C worms, and to limit $c\le0$ for $25^{\circ}$C worms. 

\subsection*{Constraints on parameters for mutants} 
\label{Section:mut_const}
The {\em ins-1} mutation suppresses the starvation avoidance behavior \cite{Kodama06}. Thus in our main analysis, we make an assumption that it does not affect any of the parameters except $\tau_a$ and $A_a$ (there is no starvation and hence no avoidance at the rearing temperature for any of the mutant worms). Since {\em ins-1(nr2091)} is a putative null mutation, we chose to model its effect as completely removing the avoidance behavior, while keeping all other parameters inferred from the N2 worm unchanged. If this assumption is correct, then no additional fitting is needed to model the {\em ins-1} worms data after we set $A_a=0$.

DAF-2 and AGE-1 are homologs of an insulin receptor tyrosine kinase and a phosphoinositide 3-kinase, respectively. Mutations in these genes are known to change the avoidance behavior \cite{Kodama06}, and hence the avoidance time scale $\tau_a$, strength $A_a$, and the bias $c$, as well as the avoidance parameters to the rearing temperature, may be changed by either mutation. Thus in our main analysis we fit these parameters to the mutant data (separately for the warm and the cold worms), while keeping the rest of the dynamical and the initial conditions parameters as inferred from the N2 worms reared at the appropriate temperature. 

Note that we do not have starvation data for any of the mutants. Hence initial conditions for avoidance at the rearing and the assaying temperatures are set to 0, and $\tau_{a,15/25}$ do not enter the model.    

\subsection*{Constructing the loss function}
\label{Section:loss}
To fit models to the data we must define a loss function, which is minimized to find the optimal parameters. Designing the loss function for physical problems is, generally, complicated since it also must avoid guiding the optimization to nonphysical regimes. The following considerations went into choosing the loss in our problem.

We use a simple mean square difference between model prediction for the worm bias at time $t$ for the worm condition $\mu$, $\hat{\Theta}_\mu(t)$, and the corresponding individual-averaged worm trajectories, $\bar{\Theta}_\mu(t)$, as a measure of the quality of fit
\begin{equation}
L_{\rm fit}=\frac{1}{2}\sum_{\mu \in M}\sum_{n=0}^{T/\Delta t} \frac{1}{v_\mu(t)} (\hat{\Theta}_\mu(t_n)-\bar{\Theta}_\mu(t_n))^2.
\label{lfit}
\end{equation}
$T=4$ h is the duration of the experiment and $\Delta t = t_n-t_{n-1}=10$ s. The condition set $\mathcal{M}$ with size $M$ consists of different starvation durations: $\{0,1,2,3,5\}$ h for $15^\circ$C, $M = 5$; and $\{0,0.5,1,2\}$ h for $25^\circ$C, $M=4$. Since each of the three mutants is well fed, $M=1$ for each of them.

\paragraph*{Removing non-biological behavior}
Some parameter combinations that minimize $L_{\rm fit}$ result in the thermotactic behavior that we consider biologically unrealistic: persistently oscillating bias at $t\gg T$. We expect the bias to converge to a constant value at long times because we are not aware of any animal that maintains non-diminishing amplitude oscillatory behavior forever. The long-term constant bias potentially differs from zero due to rearing-temperature dependent asymmetry in the thermotactic behavior \cite{Yamada03,Luo2014,Kimata2012}, and it is potentially different from the value of the temperature dependent bias $c$ in our model Eq.~\ref{ttx}. We impose this by adding the following term to the loss function
\begin{equation}
L_{\rm far}=\frac{\gamma }{2} \sum_{\mu \in \mathcal{M}} \sum_{n=0}^{T/\Delta t} \frac{M}{\sum_{\nu\in \mathcal{M}} \bar{v}_\nu} (\hat{\Theta}_\mu(t_n+T_{\rm{far}})-\bar{\hat{\Theta}})^2.
\end{equation}
This term suppresses the mean squared difference between the model predicted trajectories $\hat{\Theta}_\mu(t+T_{\rm{far}})$ starting at time $T_{\rm far}=16$ h (long after the 4 h of assaying in the droplet) and the time and the condition averaged model trajectories  $\bar{\hat{\Theta}} = \frac{\Delta t}{M\cdot T}\sum_{\mu \in \mathcal{M}} \sum_{n=0}^{T/\Delta t} \hat{\Theta}_\mu(t_n + T_{\rm{far}})$ after time $T_{\rm far}$. Here $\Delta t = 10$ s is the sampling interval, and the ratio $T/\Delta t = 4 \text{ h}/10 \text{ s} = 1440$ counts the number of sample points per trajectory. We use  $\gamma = 0.1$ as a dimensionless weight in what follows. This is about the smallest value of $\gamma$ that is able to prevent sustained oscillations in our fits. There is a wide range of $\gamma>0.1$ that similarly suppresses the oscillations, and still does not result in the degradation of the quality of fit to the experimental data.

\paragraph*{Removing flat regions in the loss function} Empirically, our optimization landscape abounds with flat regions, where the model fits have low sensitivity to large correlated parameter changes. To break this near-invariance and allow gradient-based methods to find minima faster, we penalize very small and very large parameter values with 
\begin{equation}
     L_{\rm param}=\lambda \sum_{k=1}^{K}  (\ln\theta_k)^2,
\end{equation}
where $\theta_k$, $k=1,\dots,K$ are all parameters in the model. We use $\lambda=0.1$ is what follows, which is about the smallest value of the constraint that still allows the gradient descent to converge. The quality of the fits evaluated with and without the constraint at this $\lambda$ does not change significantly.

\paragraph*{Normalization of the loss function}
In Eq.~(\ref{lfit}), $L_{\rm fit}$ scales as $T$, the total number of time points in the data. However, the data points in the experimental thermotactic trajectories are correlated because of the slow time scales of the worm behavior. To reflect that there is an effective number of independent points in the thermotactic trajectory, we must normalize $L_{\rm fit}$ to scale linearly not with the total experiment duration, but with the number of independent measurements over the duration.  For this, we compute the auto-correlation time $T_{\rm corr}$ of the residual $\hat{\Theta}_\mu(t)-\bar{\Theta}_\mu(t)$ over all conditions $\mu$ for worms raised at the specific rearing temperature. The number of independent measurements is then $n=T/T_{\rm corr}$, or, alternatively, $L_{\rm fit}$ must be rescaled by $\Delta t/T_{\rm corr}$, where $\Delta t=10$s is the temporal resolution of the experiment. The auto-correlation time $T_{\rm corr}\approx 13.3$ min for $15^{\circ} C$, and it is $T_{\rm corr}\approx 16.7$ min for $25^{\circ} C$.

\paragraph*{The overall loss function}
Overall, the considerations above yield the following loss function
\begin{dmath}
L =  \frac{\Delta t}{T_{\rm corr}}\left(\frac{1}{2}\sum_{\mu \in \mathcal{M}}\sum_{n=0}^{T/\Delta t} \frac{1}{v_\mu(t)} (\hat{\Theta}_\mu(t_n)-\bar{\Theta}_\mu(t_n))^2 + \frac{\gamma }{2} \sum_{\mu \in \mathcal{M}} \sum_{n=0}^{T/\Delta t} \frac{M}{\sum_{\nu\in \mathcal{M}} \bar{v}_\nu} (\hat{\Theta}_\mu(t_n+T_{\rm{far}})-\bar{\hat{\Theta}})^2+\lambda \sum_{k=1}^{K}  (\ln\theta_k)^2\right),\label{totalloss}
\end{dmath}
which is what we optimize to fit parameters of the model to data.
 
\subsection*{Optimization and parameter values}
\label{Section:optimization}
To fit the model to the data we vary the optimization parameters $\vec{\theta}$ in order to minimize to the loss function. For each worm type (rearing temperature and mutation), we start from $\sim 1000$ initial values of the parameters being optimized. Since some parameters take on a certain sign for biophysical reasons, we enforce these constraints by casting $\theta_i = \sign(\theta_i)\exp(\theta_i')$. This means that $\vec{\theta}'$ is unconstrained and allows us to sample the parameters $\vec{\theta}$ across many orders of magnitude. To obtain parameter values which are $\mathcal{O}(1)$, components of the vector $\vec{\theta}'$ are sampled from a uniform distribution on the unit interval.

We then use the Quasi-Newton method to minimize the loss function starting from the initial value. To determine the quality of the fit we utilize the $\chi^2$ per degree of freedom:
\begin{equation}
\chi^2/f =  \frac{L_{\rm fit}}{M f}.
\end{equation}
This is a rescaled part of the of the overall loss function, responsible only for the quality of the fit. A value $\chi^2/f \sim 1$ implies an excellent fit of the model to all data points in the fitted condition. 
 
\paragraph*{The effects of $\gamma$ on the quality of  fit} The loss function, Eq.~(\ref{totalloss}), includes the term $L_{\rm far}$, which penalizes nonphysical sustained oscillations in the thermotactic index at long times. The term is weighted by a coefficient $\gamma$, compared to the quality of fit term. The choice of $\gamma$ is unclear {\em  a priori}. In Tbl.~\ref{tab:gamma}, we report the dependence of $\chi^2/f$ on  $\gamma$. The change in the quality of fit is  $\sim 1 \%$ in response to many orders of magnitude changes in $\gamma$. We conclude that $L_{\rm far}$ is able to dampen the oscillations without significantly altering the fit quality. We thus choose $\gamma=0.1$ for all fits reported in this work, which is about the smallest value of $\gamma$ able to suppress the oscillations.

\begin{table}[]
\begin{center}
\begin{tabular}{|c||c|c|c|c|}
\hline \hline
$\gamma$ & .1    &  1    & 10  & $10^6$\\ \hline
$\chi^2/f$ & 1.17& 1.17 & 1.17 & 1.25 \\ \hline \hline
\end{tabular}
\end{center}
\caption{\label{tab:gamma} $\chi^2/f$ as a function of the dimensionless weight $\gamma$, which penalizes long-term oscillations. Quality of fit for the $15^{\circ}$C worms with all starvation durations is tabulated at $\lambda=0.1$. The impact of  $\gamma>0.1$ on the quality of fit is negligible.} 
\end{table}
 
\paragraph*{The effects of $\lambda$ on the quality of fit}  The third term of the loss function, $L_{\rm param}$, penalizes very small and very large values of parameters. It is weighted by the parameter $\lambda$ relative to the goodness of fit. In other words, the larger $\lambda$ is, the smaller the region of parameters explored in fitting. Just like for $\gamma$ above, setting the value of $\lambda$ is impossible based on first principles. We explore the dependence of $\chi^2/f$ on $\lambda$ in Tbl.~\ref{tab:lambda}. As long as $\lambda<1000$, the dependence is minimal.  We thus choose to work with the smallest value $\lambda=0.1$, which enabled an effective parameter search, for all results reported here. 
\begin{table}[]
\begin{center}
\begin{tabular}{|c||c|c|c|c|c|}
\hline \hline
$\lambda$ & .1    &  1    & 10  & 100 & 1000\\ \hline
$\chi^2/f$ & 1.17& 1.18 & 1.18 & 1.19 & 1.76\\ \hline \hline
\end{tabular}
\end{center}
\caption{\label{tab:lambda} $\chi^2/f$ as a function of the dimensionless weight $\lambda$, which penalizes large and small parameter values.  Quality of fit for the $15^{\circ}$C worms with all starvation durations is tabulated at $\gamma=0.1$. As long as $\lambda$ stays below $\sim100$, the effect on  $\chi^2/f$ is minimal.} 
\end{table}

\subsection*{Model reduction and fitted values}
\label{Section:reduction}

We choose to use fitted parameter combinations that are strongly constrained by the data as a hard algebraic constraint on the parameter values, thus reducing the number of parameters by one per constraint.  We measure the effects on such reduction on the quality of fit using the Bayesian Information Criteria (BIC) \cite{Schwarz78}. That is, the original model and all of the reduced model are assigned a Bayesian score, which, in our case, is 
\begin{equation}
BIC = k \ln(n M)+2\left(\hat{L}_{\rm fit}\frac{\Delta t}{T_{\rm corr}}\right).\label{BIC}
\end{equation}
Here, as always, $n=T/T_{\rm corr}$, and $M$ is the number of different starvation conditions. Further, $\hat{L}_{\rm fit}$ is the optimal value of the first (quality of fit) term in the loss function, Eq.~(\ref{totalloss}), $k$ is the total number of parameters in the model (including the initial condition that must be fit). The score balances the complexity of the model (the first term in Eq.~(\ref{BIC})) with the quality of the fit (the second term). According to  BIC, the reduced model is statistically better than the full, unconstrained model if its score is lower. More precisely, the posterior odds of two models are given by $P_1/P_2\approx \exp (BIC_2-BIC_1)$. 

We emphasize that there are no first-principle reasons for the model reduction. In fact, they may not correspond to realistic biophysical constraints, may not be interpretable, and are only useful to the extent that they simplify the fitting, while not decreasing the quality of the fits significantly. Thus we do not push the reduction to the extreme, and stop when the algebraic constraints coming from the reduction become uninterpretable. 

With this, for $15^\circ$C wild type worms, we identify the following candidate parameter reductions. First,  $a_{1,15}(0)=a_{0,15}(0)=0$ and $a_{5,15}(0)=a_{3,15}(0)$, so that there are only two independent initial conditions for the avoidance at the rearing temperature.  In other words, the rearing temperature avoidance for a well-fed worm and a 1 h starved worm are the same, and so are the avoidances for 3 and 5 h of starvation (the former presumably because 1 h of starvation is not enough to excite large avoidance, and the latter presumably because the avoidance gets saturated). Second, we expect $c= 0$, so that there is no long-term thermotactic bias. Third, we further verify if $\tau_{h,15}/\tau_{a,15}=\tau_h/\tau_a$. If true, this would signify that the ratio of habituation to avoidance time scales is independent of the temperature where it is measured (rearing or assaying). 

There are 17 parameters in the unconstrained model for $15^\circ$C data: four assaying and four rearing temperature avoidance initial conditions $a_s(0)$ and $a_{s,15}(0)$ for different starvation times (we remind the reader that $a_0(0)=a_{0,15}(0)=0$ for this case by construction), habituation and avoidance time scales for the memory at the assaying ($\tau_h$, $\tau_a$) and the rearing ($\tau_{h,15}$, $\tau_{a,15}$) temperatures, rearing and assay temperature habituation initial conditions ($h_{15}(0)$, $h(0)$) (both are assumed to be saturated and hence independent on the starvation duration), maximum possible values of avoidance ($A_a$) and habituation ($A_h$) at the assay temperature during assaying, and finally the long term bias $c$. The 17-parameter model has the BIC score of $BIC_{17}=176.6$. We progressively reduce the parameters one at a time and calculate the BIC scores for the reduced models. First, $c=0$ (16 parameters), we have $BIC_{16}=173.2$. We then set  $a_{1,15}(0)=a_{0,15}(0)=0$ (15 parameters) to get  $BIC_{15}=168.6$. Further, setting  $a_{5,15}(0)=a_{3,15}(0)$, we obtain $BIC_{14}=165.0$. Finally, requesting that $\tau_p/\tau_n = \tau_{p,\rm{train}}/\tau_{n,\rm{train}}$ results in $BIC_{13}=161.2$. This is the final model we use. Each of the models in the sequence is progressively more likely than the previous one. Specifically, the BIC scores give the odds $P_{13}:P_{17}\approx 2200\!:\!1$. The inferred parameters for this final model are in Tbl.~\ref{tab:N2-15}. The best fit value for this model is $\chi^2/f = 1.17$.

For $25^\circ$C wild-type worms, we identify the following putative model reductions. First, we set $a_1(0)=a_{0.5}(0)$, so that the initial conditions for the avoidance at the droplet temperature are the same for starvation durations of $0.5$ and $1.0$ h. This is reasonable since the transfer of the worm from the rearing plate to the plate with no food for starvation itself takes $\sim0.25$h, making the two starvation durations similar to each other. We then explore $A_h=A_a$, so that the maximum value of the habituation and avoidance are the same in the droplet. Finally, like for the cold worm, we try $\tau_h/\tau_a = \tau_{h,25}/\tau_{a,25}$, so that, while the actual time scales for the memories at the rearing and the experimental temperature may differ during the assaying in the droplet, the ratios of habituation and avoidance time scales stay the same. 

The $25^\circ$C model starts with 15 parameters (we have one fewer starvation duration than for the cold worm, and hence two fewer initial conditions). For this model, $BIC_{15}=123.6$. Setting  $a_1(0)=a_{0.5}(0)$ results in $BIC_{14}=120.2$. Adding  $A_h=A_a$ gives $BIC_{13}=116.6$. Finally, with  $\tau_h/\tau_a = \tau_{h,25}/\tau_{a,25}$, we have $BIC_{12}=113.2$. Again, every next model in the sequence is more likely than the previous one, and the odds $P_{12}:P_{15}\approx 181:1$.  The inferred parameters for this final model are in Tbl.~\ref{tab:N2-25}. The best fit value for this model is $\chi^2/f = 1.12$.

\begin{table}[]
\begin{center}
\begin{adjustbox}{width=\textwidth,center=\textwidth} 

\begin{tabular}{|c||c|c|c|c|c|c|c|c|c|c|}
\hline \hline
 Parameter       & $\tau_h/\tau_a$ & $A_h$ & $h(0)$ & $\tau_a$ & $A_{a}$ & $a_{0}(0)^*$ & $a_{1}(0)$ & $a_{2}(0)$ & $a_{3}(0)$ & $a_{5}(0)$  \\ \hline \hline
 value           &  0.28           & 6.37  & -1.92  & 1.39     & 6.44    & 0            & -1.68      & -1.86      &  -2.02     & -2.91        \\ \hline
$\Sigma_{ii}$    & 0.14            & 2.3   & 0.88   & 0.58     & 5.1     &              & 0.84       & 0.84       &   0.85     & 1.2          \\ \hline
 $h_{ii}^{-1/2}$ & 0.009           & 0.2   & 0.08   & 0.06     & 0.2     &              & 0.06       & 0.08       &   0.2      & 0.8          \\ \hline \hline
 Parameter       & $\tau_{h,15}/\tau_{a,15}$ & $h_{15}(0)$ & $\tau_{a,15}$ & $a_{0,15}(0)^*$ & $a_{1,15}(0)$  & $a_{2,15}(0)$  & $a_{3,15}(0)$ & $a_{5,15}(0)$ & $c$ & $\Theta_0^*$ \\ \hline \hline
 value           & $\tau_h/\tau_a$           & -1.15       & 6.66          & 0               & $a_{0,15}(0)$  & -0.51          & -1.41         & $a_{3,15}(0)$ & 0   &  0.27        \\ \hline
 $\Sigma_{ii}$   &                           & 0.48        & 4.9           &                 &                & 0.42           & 0.84          &               &     &              \\ \hline
 $h_{ii}^{-1/2}$ &                           & 0.05        & 0.4           &                 &                & 0.09           & 0.2           &               &     &              \\ \hline \hline
\end{tabular}
\end{adjustbox}
\end{center}
\caption{\label{tab:N2-15}Optimal parameter values and their uncertainties (quantified by $\Sigma$ and $h$, see text) for the final, 13-parameter model, describing the thermotactic dynamics of the cold N2 worm (reared at $15^{\circ}$C). BIC favors this model over a full 17-parameter model with odds $\approx2200:1$. The quality of the fit is $\chi^2/f = 1.17$.  $\Sigma_{ii}$ are the estimates of the upper bound on the parameter uncertainty, accounting for variation of the other parameters, and $h_{ii}^{1/2}$ are the estimates of the lower bound on the uncertainty; see text for details. Parameters indicated by $^*$ are set a priori and are not fitted. Parameters with values relating them to other parameters are set by the model reduction. Parameter values are defined in Eqs.~(\ref{eqp}-\ref{ttx}). Briefly:  $\tau_h/\tau_a$ -- ratio of time scales for habituation and avoidance dynamics at the droplet temperature during the droplet assay; $A_h$ -- the maximum possible value of habituation at the droplet temperature during the assay; $h(0)$ -- the initial value of the habituation at the droplet temperature at the beginning of the assay (hours); $\tau_a$ -- the time scale of the avoidance at the droplet temperature during the assay; $A_a$ -- the maximum possible value of the avoidance at the droplet temperature during the assay; $a_s(0)$ -- initial values of the avoidance at the droplet temperature at the beginning of the assay for starvation durations $s$; $\tau_{h,15}/\tau_{a,15}$ -- the ratio of the time scales for the habituation and avoidance at the rearing temperature of 15$^\circ$C during the assay; $h_{15}(0)$ -- initial value of the habituation at the rearing temperature at the beginning of the assay; $a_{s,15}(0)$ -- initial values of the avoidance at the rearing temperature at the beginning of the assay for starvation durations $s$; $c$ -- thermotactic bias; $\Theta_0$ -- saturating value of the worm bias. }
\end{table}


\begin{table}[]
\begin{center}
\begin{adjustbox}{width=\textwidth,center=\textwidth} 

\begin{tabular}{|c||c|c|c|c|c|c|c|c|c|}
\hline \hline
 Parameter       & $\tau_h/\tau_a$ & $A_h$ & $h(0)$ & $\tau_a$ & $A_{a}$ & $a_{0}(0)^*$ & $a_{0.5}(0)$ & $a_{1}(0)$   & $a_{2}(0)$   \\ \hline \hline
 value           &  0.41           & $A_a$ &  0.25  & 0.75     & 7.37    & 0            & 0.35.        & $a_{0.5}(0)$ &  0.68                 \\ \hline
$\Sigma_{ii}$    & 0.29            &       &  0.27  & 0.39     & 5.2     &              & 0.28         &              &   0.46                    \\ \hline
 $h_{ii}^{-1/2}$ & 0.02            &       &  0.1  & 0.06     & 0.6     &              & 0.1          &              &   0.3                  \\ \hline \hline
 Parameter       & $\tau_{h,25}/\tau_{a,25}$ & $h_{25}(0)$ & $\tau_{a,25}$ & $a_{0,25}(0)^*$ & $a_{0.5,25}(0)$ & $a_{1,25}(0)$  & $a_{2,25}(0)$ & $c$    & $\Theta_0^*$ \\ \hline \hline
 value           & $\tau_h/\tau_a$           & 0.68        & 2.66          & 0               & 0.63            &  1.2           &  1.42         & -0.1   &  0.27        \\ \hline
 $\Sigma_{ii}$   &                           & 0.36        & 1.4           &                 & 0.37            & 0.49           &  0.66         &  0.14  &              \\ \hline
 $h_{ii}^{-1/2}$ &                           & 0.09        & 0.4           &                 & 0.1             & 0.2            &  0.2          &  0.03  &              \\ \hline \hline
\end{tabular}
\end{adjustbox}
\end{center}
\caption{\label{tab:N2-25}Optimal parameter values for the final, 12-parameter model, describing the thermotactic dynamics of the warm N2 worm (reared at $25^{\circ}$C). BIC favors this model over a full 15 parameter model with odds $\approx181\!:\!1$. The quality of the fit is $\chi^2/f = 1.12$. Notation used is the same as in Tbl.~\ref{tab:N2-15}.}
\end{table}

\subsection*{Mutant fits}
\label{Section:Mutant_fits}
As explained in \nameref{Section:mut_const}, we inherit most of the parameters for the mutants from the N2 fits. We do not need to fit any additional parameters for {\em ins-1} mutants. For {\em age-1} and {\em daf-2}, we only fit the avoidance parameters $\tau_a$ and $A_a$, as well as the long-term bias $c$. While we try to keep $c=0$ for the cold worms, as for N2, the obtained fit for the 15$^\circ$C {\em daf-2} worm is bad ($\chi^2/f>2)$, and so $c\neq0$ is also allowed here.  The fitted values of the parameters and the quality of fits are in Tbl.~\ref{tab:mut-avoidance}.

We additionally argue that fitting just the habituation (rather than the avoidance) parameters does not produce good fits, so that mutations, indeed, affect the avoidance and not the habituation pathway. The fitted values of the parameters and the quality of fit for this model are in Tbl.~\ref{tab:mut-habituation}.

\begin{table}[!ht]
    \centering
    \begin{tabular}{|c||c||c|c|c||c|c|c|}
    \hline \hline
            \multicolumn{2}{|c||}{Temperature}    &\multicolumn{3}{c||}{$15^\circ C$}  & \multicolumn{3}{c|}{$25^\circ C$}    \\  \hline  \hline
                                & Parameter       & $\tau_a$     & $A_a$ & $c$       & $\tau_a$     & $A_a$   & $c$            \\  \hhline{~=======}
        ~                       & value           & 0.39         & 0     & 0         & 6.9          & 0       & 1.5            \\  \cline{2-8}
        {\it ins\/}-1           & $\Sigma_{ii}$   & ~            & ~     & ~         & ~            & ~       & ~              \\  \cline{2-8}
        ~                       & $h_{ii}^{-1/2}$ & ~            & ~     & ~         & ~            & ~       & ~              \\  \hline  \hline
                                & Parameter       & $\tau_a$     & $A_a$ & $c$       & $\tau_a^*$   & $A_a^*$ & $c^\dagger$    \\  \hhline{~=======}
                                & value           & 7.5          & 156   & -9.7      & 1.2          & 3.3     & 0.026          \\  \cline{2-8}
      {\it daf\/}-2             & $\Sigma_{ii}$   & 6.6          & 260   & 11        & 0.64         & 1.1     & 0.19           \\  \cline{2-8}
                                & $h_{ii}^{-1/2}$ & 0.27         & 4.7   & 0.34      & 0.53         & 0.28    & 0.046          \\  \hline \hline
                                & Parameter       & $\tau_a$     & $A_a$ & $c$       & $\tau_a$     & $A_a$   & $c$            \\  \hhline{~=======}
        ~                       & value           & 3.4          & 35    & 0         & 6.9          & 120     & 1.5            \\  \cline{2-8}
        {\it age\/}-1           & $\Sigma_{ii}$   & 1.6          & 16    & ~         & 9.0          & 170     & 0.78           \\  \cline{2-8}
        ~                       & $h_{ii}^{-1/2}$ & 0.08         & 0.23  & ~         & 0.32         & 5.5     & 0.22           \\  \hline  \hline
    \end{tabular}
    \caption{\label{tab:mut-avoidance}Optimal parameter values and their uncertainties for the parameters that change between the N2 worm and the mutant worms ({\it daf}-2 and {\it age}-1) in the model that assumes that the mutations affect the avoidance pathway only. There are 3 or 2 such parameters depending on the mutant and the rearing temperature. All notations are as in Tbl.~\ref{tab:N2-15}. The quality of fit values are listed in Tbl.~\ref{tab:mut-fitvals}.  Note that increases (decreases) in the parameter indexed by $*$ ($\dagger$) within $+h_{ii}^{-1/2}$ ($-h_{ii}^{-1/2}$) result in nonphysical long-term sustained oscillations in the model.  The {\it ins}-1 mutant is not fitted, instead $A_a$ is set to zero while $\tau_a$ and $c$ are inherited from Tbls.~\ref{tab:N2-15}, and  ~\ref{tab:N2-25}.} 
\end{table}

\begin{table}[!ht]
    \centering
    \begin{tabular}{|c||c||c|c|c||c|c|c|}
    \hline \hline
            \multicolumn{2}{|c||}{Temperature}    &\multicolumn{3}{c||}{$15^\circ C$}  & \multicolumn{3}{c|}{$25^\circ C$}    \\  \hline  \hline
                                & Parameter       & $\tau_h$     & $A_h$ & $c$         & $\tau_h$     & $A_h$      & $c$      \\ \hhline{~=======}
        ~                       & value           & 113.9        & 158.1 & 0           & 0.59         & 10.9       & -0.1     \\  \cline{2-8}
        {\it ins\/}-1           & $\Sigma_{ii}$   & 19.9        & 20.1   & ~           & 0.17         & 0.87       & ~       \\  \cline{2-8}
        ~                       & $h_{ii}^{-1/2}$ & 19.8         & 20.0  & ~           & 0.12         & 0.59       & ~      \\  \hline  \hline
                                & Parameter       & $\tau_h$     & $A_h$ & $c$         & $\tau_h$     & $A_h$      & $c$        \\  \hhline{~=======}
                                & value           & 0.11         & 5.5   & -0.5        & 0.39         & 11.8       & -0.014          \\  \cline{2-8} 
      {\it daf\/}-2             & $\Sigma_{ii}$   & 0.04         & 0.14  & 0.07        & 0.16         & 0.86       & 0.12           \\  \cline{2-8}
                                & $h_{ii}^{-1/2}$ & 0.02         & 0.14  & 0.03        & 0.14         & 0.51       & 0.08          \\  \hline \hline
                                & Parameter       & $\tau_h$     & $A_h$ & $c$         & $\tau_h$     & $A_h$      & $c$        \\  \hhline{~=======}
        ~                       & value           & 0.06         & 3.8   & 0           & 0.12         & 5.4        & 0.08            \\  \cline{2-8}
        {\it age\/}-1           & $\Sigma_{ii}$   & 0.03         & 0.30  & ~           & 0.01         & 0.20       & 0.09           \\  \cline{2-8}
        ~                       & $h_{ii}^{-1/2}$ & 0.02         & 0.25  & ~           & 0.01         & 0.18       & 0.08           \\  \hline  \hline
    \end{tabular}
  \caption{\label{tab:mut-habituation}Optimal parameter values and their uncertainties for the parameters that change between the N2 worm and the mutant worms ({\it daf-2} , {\it age-1}, and {\em ins-1}) in an alternative model, where the mutations affect the habituation pathway only. All notations are as in Tbl.~\ref{tab:N2-15}. The quality of fit values are listed in Tbl.~\ref{tab:mut-fitvals}. As the data in Tbl.~\ref{tab:mut-fitvals} shows, the fits overall are worse than for our primary model.}. 
\end{table}

\begin{table}[!ht]
    \centering
    \begin{tabular}{|c||c|c||c|c|}
    \hline \hline
     Temperature        &\multicolumn{2}{c||}{Avoidance}  & \multicolumn{2}{c|}{Habituation}    \\  \hline  \hline
      $15^{\circ}$C     & $\chi^2/f$   & BIC   & $\chi^2/f$  & BIC                  \\  \hhline{=====}
      {\it ins\/}-1     & 1.28         & 23.1  & 1.19        & 27.2                 \\   \cline{1-5} 
      {\it daf\/}-2     & 1.26         & 31.4  & 1.76        & 40.4                  \\  \cline{1-5}
      {\it age\/}-1     & 1.20         & 27.4  & 2.37        & 48.4                 \\   \hline \hline

      $25^{\circ}$C     & $\chi^2/f$   & BIC   & $\chi^2/f$  & BIC                  \\  \hline \hline
      {\it ins\/}-1     & 1.59         & 22.9  & 0.89        & 18.2                 \\  \cline{1-5}
      {\it daf\/}-2     & 1.06         & 23.3  & 1.06        & 25.1                  \\  \cline{1-5}
      {\it age\/}-1     & 1.98         & 36.5  & 2.36        & 42.0                  \\  \hline  \hline
      Total BIC         & ~            & 164.5 & ~           & 201.3             \\  \hline  \hline
    \end{tabular}
  \caption{\label{tab:mut-fitvals} $\chi^2$  per degree of freedom and BIC scores  for the model fits appearing in Tbl.~\ref{tab:mut-avoidance} and  \ref{tab:mut-habituation}. Overall, the BIC scores imply that the model where mutations affect the avoidance pathway rather than the habituation pathway  explains the data much better,  with the odds of $9.6\times 10^{15}\!:\!1$ in its favor.} 
\end{table}

\subsection*{Parameter and model trajectory error bars }
\label{Section:error_bars}
In Tbls.~\ref{tab:N2-15} and \ref{tab:N2-25}, we report the best fit parameter values for the reduced models for the cold and the warm worms. Since, even after the model reduction, there are still many parameter combinations that result in similar dynamics, we need to quantify the uncertainty on both the fitted parameter values, and on the dynamics themselves. Due to the near-degeneracy of the loss function for different parameter values, we consider the uncertainty on the dynamics a more important characteristic of the model fit than the parameter uncertainty.  

\paragraph*{Parameter error bars}
\label{Section:bands}

For estimating parameter uncertainties, we make the usual assumption that the loss function is quadratic in the parameter values around the optimum, $\vec{\theta}^*$. While not strictly true, the inaccuracies introduced by this assumption are not critical, precisely because we consider the trajectories, and not the parameter values, as the important properties of the model fits. The Hessian around the optimal value $H_{ij} = \partial_{\theta_i}\partial_{\theta_j}L |_{\vec{\theta}^*}$ is estimated for us during the optimization by MATLAB's {\tt fminunc()} using the finite differences method.

Parameters have correlated effects on model fits. Thus we report two measures of their uncertainty. The first is a lower bound on the error bar of each $\theta_i$, obtained as $1/\sqrt{H_{ii}}$. This quantity measures how much the parameter can change and not affect the loss function significantly, while keeping all other parameters fixed. The second is the upper bound on the uncertainty in the quadratic approximation, $\Sigma_{ii}=\left(H^{-1}|_{\theta^*}\right)_{ii}$. This quantity measures how much a specific parameter can change without affecting the quality of the fit, while allowing variability in other parameters to compensate for the effects of changes in the explored parameter.  Both errors are shown for the reduced model of the N2 worms in Tbls.~\ref{tab:N2-15} and \ref{tab:N2-25}, and for the mutant worms in Tbls.~\ref{tab:mut-avoidance} and \ref{tab:mut-habituation}.

\paragraph*{Wildtype trajectory error bands}
To estimate the uncertainty in the fitted dynamics, we use bootstrapping \cite{Hesterberg2011}. We use a single worm (rather than a single time point in a worm's dynamics) as a single unit of data \cite{Saravanan2020}. We sample with replacement $N_\mu$ worms from $N_\mu$ worms in the experimental condition $\mu$ (here $\mu$ defines the rearing temperature and the starvation duration). From the resampled data, for each condition, we calculate the new four-hour average thermotactic index curve $\bar{\Theta}_\mu(t)$. We repeat this a 100 times, thus obtaining $\{\bar\Theta_{\mu, k}(t)\}, k=1,\dots,100$. Each of the bootstrapped trajectories is fit by the model, Eqs.~(\ref{eqp}-\ref{ttx}), producing parameter fits $\{\hat{\vec{\theta}}_{k}\}$ and obtaining predictions $\{\hat\Theta_{\mu, k}(t)\}, k=1,\dots,100$. For each time and condition, we report the error bands as the $16.5\dots83.5$ range within the set $\{\hat\Theta_{\mu, k}(t)\}$. These bands would correspond to a one standard deviation confidence band if the statistics of the trajectories were Gaussian (which they are not). 

\paragraph*{Mutant trajectory error bands}
Mutants reared at warm or cold temperature inherit most parameters from their respective N2 worms, except for a handful parameters (those describing avoidance in the primary model, and habituation in the alternative model), which are fitted to the mutant data. Thus they inherit the bootstrapped N2 data as well, and the associated best fit parameters,  $\{\hat{\vec{\theta}}_{k}\}$, and model predictions, $\{\bar\Theta_{\mu, k}(t)\}, k=1,\dots,100$. If additional mutant-specific parameters are fitted, for such mutants, we generate the bootstrapped data with 100 sets of worms resampled with replacement, similarly to the wildtype worms. We then pair, at random, each mutant resampled data set with one of the N2 resampled data sets, inherited from estimating uncertainty of the N2 models. For each such pair, we use its fitted N2 parameters and fit the mutant parameters with these N2 ones. The 16.5 to 83.5 percentile range of the fits is then reported as the error band.  

\bibliographystyle{Science}
\bibliography{scibib.bib}
\section*{Acknowledgements}
We would like to thank Greg Stephens, Fred Bartumeus,  C.\ Randy Gallistel, and Gordon Berman for useful conversations, and the Aspen Center for Physics for hospitality.
\subsection*{Funding:} We acknowledge the support of the Natural Sciences and Engineering Research Council of Canada (NSERC), Human Frontier Science Program (HFSP), U.S.\ National Science Foundation (NSF), and the Simons Foundation.
\subsection*{Author contributions:} W.S.R.\ and I.N.\ conceived and designed the project; K.P.\ performed all of the experiments; A.R.\ and K.P.\ performed the data analysis; A.R.\ and I.N.\ developed and fitted the model;
A.R., I.N., and W.S.R.\ wrote the manuscript. W.S.R.\ and I.N.\ supervised the study.
\subsection*{Competing interests:} None.
\subsection*{Data and material availability:} Thermotactic index data is available at \url{https://figshare.com/articles/dataset/C_elegans_Data_zip/19907110}


\end{document}